\documentclass[aps,pra,superscriptaddress]{revtex4}
\pdfoutput=1
\usepackage[intlimits]{amsmath}
\usepackage{amsfonts}
\usepackage{psfrag}
\usepackage{enumerate}
\usepackage{subfigure}
\usepackage{enumitem}
\usepackage[usenames]{color}
\advance\voffset by -1cm
\advance\hoffset by -0.5cm
\newcommand\be            {\begin{equation}}
\newcommand\ee            {\end{equation}}
\newcommand\bes           {\begin{subequations}}
\newcommand\esu           {\end{subequations}}

\newcommand{\ea}[1]{\begin{align}#1\end{align}}

\newcommand{\bigx}[1]{\bBigg@{#1}}

\def\3pt#1#2#3{{\langle{#1}\vert{#2}\vert{#3}\rangle}}

\newcommand\doi[2]        {\href{http://dx.doi.org/#1}{#2}}

\newcommand{\EQ}{\begin{equation}}
\newcommand{\EN}{\end{equation}}
\usepackage{epsfig}
\usepackage{color}
\usepackage{psfrag}
\usepackage{amsmath}
\usepackage{graphicx}
\usepackage{amsfonts}
\usepackage{amssymb}
\begin{document}
\bibliographystyle{plainnat}

\title{{\Large {\bf Quench Dynamics in Two-Dimensional\\
Integrable SUSY Models}}}

\author{Axel Cort\'es Cubero}
\affiliation{SISSA and INFN, Sezione di Trieste, via Bonomea 265, I-34136, 
Trieste, Italy}

\author{Giuseppe Mussardo}
\affiliation{SISSA and INFN, Sezione di Trieste, via Bonomea 265, I-34136, 
Trieste, Italy}
\affiliation{International Centre for Theoretical Physics (ICTP), 
I-34151, Trieste, Italy}

\author{Mi{\l}osz Panfil}
\affiliation{SISSA and INFN, Sezione di Trieste, via Bonomea 265, I-34136, 
Trieste, Italy}

\begin{abstract}
\noindent
We analyse quench processes in two dimensional quantum field theories with infinite number of conservation laws which also include fermionic 
charges that close a $N=1$ supersymmetric algebra. While in general the quench protocol induces a breaking of supersymmetry, we show that there are particular initial states which ensure the persistence of supersymmetry also for the dynamics out of equilibrium. We discuss the conditions that identify such states and, as application, we present the significant cases of the Tricritical Ising Model and the Sine-Gordon model at its supersymmetric point. We also address the issue of the Generalized Gibbs Ensemble in the presence of fermionic conserved charges.

\vspace{3mm}
\noindent
Pacs numbers: 11.10.Wx, 11.55.Ds, 11.30.Pb

\end{abstract}
\maketitle


\section{Introduction}

A problem that has recently attracted a lot of attention -- both from the theoretical and experimental points of view -- consists of  the dynamics out of equilibrium of an extended quantum system prepared at $t=0$ in a state $\mid B\, \rangle$ that is not an eigenstate of its Hamiltonian $H$ (see, for instance, \cite{Weiss,Deutsch,CC,SilvaReview,IC,Berges,Rigol,FM,EMP,CEF,GGErecent,Bertin}). In many cases of physical interest,  such a state $\mid B \, \rangle$ -- also known as  {\em boundary state} -- may be thought as the ground state of another Hamiltonian $H_0$. Hence, the dynamics out of equilibrium is essentially a {\em global quantum quench process} induced by an abrupt change of the Hamiltonians $H_0 \rightarrow H$ that occurs at $t=0$: after this time and for all $t >0$, the system evolves unitarily under $H$, i.e. $\mid B(t) \,\rangle = e^{-i H t} \mid B \, \rangle$, and for all the models put out of equilibrium in this way, the dynamical data are entirely encoded into the expectation values 
\be
G_{ab\ldots s}(x_1,t_1;  \ldots  x_n, t_n \mid t) \,=\,\frac{
\langle B(t) \mid \Omega_a(x_1,t_1) \Omega_b(x_2,t_2) \ldots \Omega_s(x_n,t_n) \mid B(t) \rangle}{\langle B \mid B \rangle}
\,\,\,,
\label{fundamentalmatrixelements}
\ee
where $\Omega_r(x_i,t_i)$ are local operators of the theory. In the following we will consider quantum systems which admit a continuum limit described by a d-dimensional relativistic Quantum Field Theory (QFT), namely a theory made of particle excitations $\mid A_{\{k\}}(p) \rangle$ labelled by: 
(i) the momentum $\vec{p}$ entering together with $p_0$ the d-dimensional dispersion relation $p^\mu p_\mu \,=\,m_k^2$, where $m_k$ is the mass of the relative particle;  (ii) a series of indices $\{k\}$ which specify, for instance, whether the particle is boson or fermion, or other possible internal degrees of freedom. Relativistic quantum field theories are known to describe scaling limit of many interesting many body theories of cold atomic gases or magnetic materials, in particular in d=2: apart from the familiar quantum Ising model -- deeply related to Majorana fermions -- it is also worth mentioning for instance the Tricritical Ising model for magnetic alloys -- whose different phases are associated to different field theories, such $\phi^6$ Landau Ginzburg, $E_7$ Toda Field Theory, Supersymmetric Wess-Zumino model, etc. -- or several strongly correlated systems, either spin chains or cold atom models, described by the Sine or Sinh-Gordon model, see for instance \cite{GMbook,Tsvelikbook,KMT}. 

In this paper, in particular,  we are interested in studying QFT invariant under supersymmetry (SUSY) transformations: these are theories which have excitations both of fermionic and bosonic type, and which are left invariant by transformations which map the ones in the others. SUSY has a very long history in physics, in particular in high-energy and mathematical physics, for its promising role in addressing long standing problems such as cancellation of infinities in perturbative expansion, the particle hierarchy problem in the Standard Model, the possibility of setting up grand-unification theories, or the mathematical consistency of string theories (see, for instance, \cite{WZ,Sonhius,Weinberg,Polchinski,mirror,OliveWitten} and references therein). In statistical or condensed matter physics, SUSY often emerges as an effective symmetry of the low-energy limit of several systems, some of which will be discussed in some detail later. In the following we will focus our attention exclusively on continuum models noting however that a substantial effort has been also devoted to studying and developing lattice supersymmetry \cite{Kareljan}. 

Our main concern is to understand under which conditions a supersymmetric QFT has a quench dynamics which is also supersymmetric. As we are going to see, the question has a certain number of subtilities, given that SUSY has features which differ from the usual internal symmetries of a field theory. In this respect, in order to understand whether SUSY is always broken once brought away from equilibrium and, if so, what is the pattern and the consequences of this breaking, we will find useful to compare the dynamics out of equilibrium with the finite temperature equilibrium situation, given the very similar geometrical formulation of both cases.  Not excluded that a dynamics of equilibrium SUSY invariant may also occur in some higher dimensional theories, in this paper however we will point out that the natural theories where to look for SUSY dynamics are those defined in $d=2$, in particular those having an infinite number of conserved currents, i.e. the supersymmetric integrable models. Two-dimensional theories have their own 
peculiarities and reasons of interest: for instance, in $d=2$ the usual distinction between fermion and boson loses its meaning, which implies that for this dimensionality of the space-time there may also be irreducible representations of the SUSY algebra that involve ordinary soliton  states, and not just bosons and fermions. This is the case, for instance, of the Sine-Gordon model at a special value of its coupling constant or the Tricritical Ising Model along its first order critical line. In this paper, for simplicity, we will focus our attention only on $N=1$ SUSY theories, where $N$ stands for the number of fermionic charges, and we will find conditions on the boundary state $\mid B \, \rangle$ such that the quench process preserves special combination of the SUSY charges. We will also discuss and comment the Generalised Gibbs Ensemble associated to $d=2$ supersymmetric theories. For completeness, it is worth pointing out that there are in literature some previous studies on SUSY theories out of equilibrium, which use however different approaches from ours \cite{devega,3dsusy}.    

The paper is organised as follows. In Section \ref{section2}, we initially identify the building blocks of the dynamics out of equilibrium. This analysis is instrumental to present, in Section \ref{section3}, the main questions which challenge the realisation of an exact SUSY invariance for the 
dynamics out of equilibrium. Section \ref{moreond=2} is devoted to the formalism of $N=1$ SUSY in $d=2$, while in Section \ref{FiniteTemperature} we consider one of the crucial questions of the SUSY dynamics out of equilibrium, alias how one can have a spontaneously symmetry breaking of SUSY at finite energy density without the appearance of the Goldstino. Section \ref{ElasticSUSYSMatrix} recalls the basic results of the exact SUSY $S$-matrix, showing in particular the underlying SUSY invariance of ordinary kink-like theories such as the Sine-Gordon model (at special value of its coupling constant) and the Tricritical Ising Model along its first order phase transition line. These results will be conducive for arguing in Section \ref{Bose&Fermion in d=2} that bosonic and fermionic occupation numbers can be equal in interacting $d=2$ theories -- a result that, together with 
the absence of Goldstino -- stresses the possibility to have a dynamics out of equilibrium which is SUSY invariant. Section \ref{SUSYBOUNDARY} is the most important part of the paper where, for integrable SUSY theories in $d=2$, we identify a set of sufficient condition that the boundary state $| B \rangle$ must satisfy in order to have a SUSY dynamics. These results are further discussed for the Sine-Gordon and the Tricritical Ising Modes in Sections \ref{SGBoundary} and \ref{TIMBoundary}, respectively.  Section \ref{SUSYsignature} discusses how SUSY helps in interpreting relations between correlation functions of different order parameters while Section \ref{SUSYGGE} faces the interesting implementation of the Generalized Gibbs Ensemble in SUSY integrable theories. Our conclusions and future perspectives can be found in the final Section  
\ref{conclusions}.

\section{Building blocks for the out of equilibrium dynamics}\label{section2}
Let's initially discuss what are the building blocks out of which one can recover, in principle, the entire non-equilibrium dynamics of a QFT following a quantum quench. As shown below, these building blocks can be identified with: (i) the particle basis in the Hilbert space; (ii) the boundary state and its particle content; (iii) the matrix elements of the local operators on the particle basis. Let's address each issue separately.   

 \vspace{1mm}
{\bf Particle Basis.} The first fundamental object one should know in order to control the dynamics out of equilibrium is the basis of the Hilbert space: in a QFT this consists of the single particle excitations $\mid A_{\{k\}}(p) \rangle$ together with all the infinitely many higher excited multi-particle states $|A_{k_1}(p_1) \ldots A_{k_n}(p_n) \rangle$. These states are chosen to be eigenstates both of the Hamiltonian $H$ and the momentum operator $\vec{P}$ 
\begin{eqnarray}
H \, \mid A_{k_1}(p_1) \ldots A_{k_n}(p_n) \rangle & \,=\, & \left(\sum_{i=1}^n E(p_i)\right) \, 
\mid A_{k_1}(p_1) \ldots A_{k_n}(p_n) \rangle \label{eigenstates1}\,\,\,,
\\
\vec{P} \, \mid A_{k_1}(p_1) \ldots A_{k_n}(p_n) \rangle & \,=\, & \left(\sum_{i=1}^n \vec{p}_i\right) \, 
\mid A_{k_1}(p_1) \ldots A_{k_n}(p_n) \rangle \label{eigenstates2} \,\,\,. 
\end{eqnarray}
Using the concise notation $\mid A_{k_1}(p_1) \ldots A_{k_n}(p_n) \rangle \,\equiv \, \mid n, \alpha \rangle$, where the Greek letter $\alpha$ 
stands for the whole collection of indices including the momenta $p_i$, these states must fulfil the completeness relation 
\be
\sum_{n=0}^\infty \,\int d\alpha\, \mid n,\alpha\rangle \langle n, \alpha \mid \,=\, 1 \,\,\,,
\label{completeness}
\ee
and the normalization condition 
\be
\langle n,\alpha \mid m,\beta \rangle \,=\,\delta_{n,m} \delta(\alpha-\beta) \,\,\,.
\label{normalization}
\ee

\vspace{1mm}
{\bf Boundary State.} 
\begin{figure}[t]
\psfig{figure=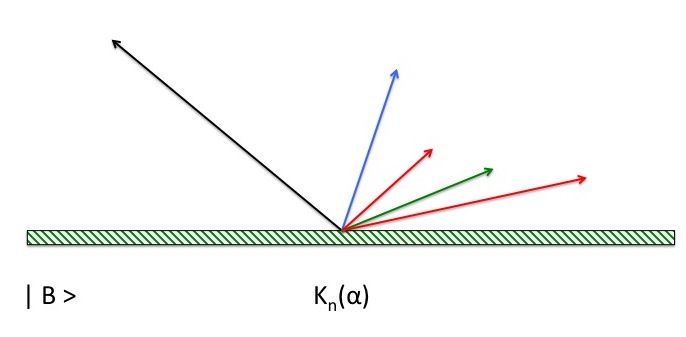,height=5cm,width=9cm}
\caption{{\em Virtual emission of $n$ particles out of the initial state $\mid B \rangle$. The color of the lines refers to the 
different internal quantum numbers of the various particles while the length of the arrows is proportional to their momenta.}}
\label{virtualemission}
\end{figure}
Since the initial state $\mid B\, \rangle$ belongs to the Hilbert space spanned by the multi-particle states, it admits an expansion on such a basis 
\be
\mid B \,\rangle \,=\, \sum_{n=0}^\infty \,\int d\alpha\,  K_n(\alpha) \, | n, \alpha \rangle \,\,\,,
\ee
where the coefficients $K_n(\alpha) \equiv \langle n, \alpha | B \rangle$ can be regarded as the amplitudes relative to the virtual emission 
at $t=0$ of $n$-particle states with quantum numbers $\alpha$ out of the boundary $|B \rangle$, as shown in Figure \ref{virtualemission}. 
In virtue of general requirements that a QFT must fulfil, these amplitudes satisfy a certain number of constraints, some of them quite simple, other more elaborated: for instance, if the system is translation invariant, the space component $\vec{P}$ of the total momentum is conserved and therefore, without losing any generality, we can take the initial state $\mid B \rangle$ to be eigenstate of $\vec{P}$ with zero eigenvalue (any other eigenvalue different from zero is in fact equivalent to a Lorentz boost transformation of the system): this implies that the amplitudes $K_n(\alpha)$ have to be proportional to $\delta\left(\vec{p_1} + \vec{p_2} + \cdots \vec{p_n}\right)$.  A more elaborated set of constraints on $K_n(\alpha)$ come from the requirement that the matrix element $\langle B \mid e^{-R H} \mid B \rangle$, defined on a finite volume $V = L^{d-1}$, scales for large $L$ as 
\be
\frac{\langle B \mid e^{-R H} \mid B \rangle}{\langle B \mid B \rangle} 
\equiv e^{- V {\mathcal F}(R)} \,\,\,.  
\label{partitionfunctionBB}
\ee
\begin{figure}[t]
\label{figurageometry}
\begin{center}
\includegraphics[width= 0.7\columnwidth]{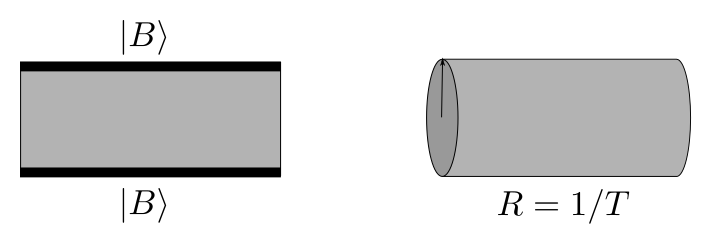}
\caption{{\em Geometry of the QFT space-time for the out-of-equilibrium (left hand side) and finite temperature system (right hand side).}}
\end{center}
\end{figure}

\noindent
The origin of this condition stays in the observation that the left hand side of this equation is nothing but the partition function $Z_{BB} \equiv e^{-F(R,V)}$ in a slab geometry of volume $V$ and width $R$, with boundary conditions at both sides of the slab set by the boundary state $|B \, \rangle$: for the extensive nature of the free-energy $F(R,V)$, this quantity must be proportional to the volume $V$, namely $F(R,V) = V {\mathcal F}(R)$, and this leads to the validity of eq.\,(\ref{partitionfunctionBB}).  In light of this exponential behaviour in $V$ of the matrix element ratio $\langle B \mid e^{-R H} \mid B \rangle /\langle B \mid B \rangle$, the boundary state $| B \,\rangle$ must then necessarily be made of an infinite number of particles and moreover the amplitudes $K_n(\alpha)$ must satisfy a series of integral equations which, although they do not fix $K_n(\alpha)$ uniquely, provide nevertheless some constraints on their behaviour \cite{FM}. Notice that eq. (\ref{partitionfunctionBB}) implies that the post-quench systems has a finite energy density for unit volume given by 
\be
\epsilon_0 \,=\,\frac{1}{V} \frac{\langle B \mid H \mid B \rangle}{\langle B \mid B \rangle} \,=\,{\mathcal F}'(0) \,\,\,.
\label{finitedensity}
\ee
For this reason, the dynamics out of the equilibrium associated to a quantum quench shares some similarities with the finite temperature equilibrium situation, since both have a finite energy density per unit volume. From a geometrical point of view, both quantum quench and finite T situations involve solving the QFT in a finite geometry (see Figure 2), where quench dynamics and finite temperature refer to a slab and cylinder geometry, respectively. This observation proves to be useful when we will discuss supersymmetric field theories out of equilibrium. Additional properties of the boundary states 
$| B\,\rangle$ and particular classes thereof, especially in relation with integrable quantum field theories,  will be discussed later. 

\begin{figure}[b]
\psfig{figure=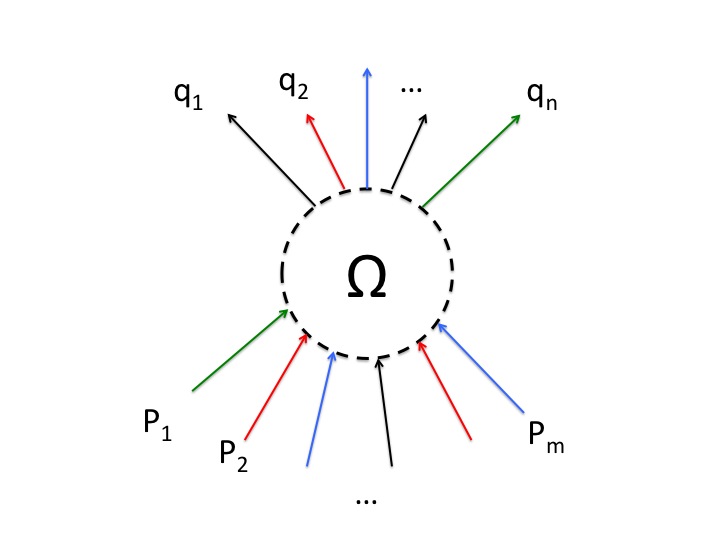,height=7cm,width=9cm}
\caption{{\em Form Factor of a local operator $\Omega(0,0)$, with $n$ in-particles and $m$ out-particles. Different colours refer to different internal quantum numbers of the particles.}}
\label{FormFactorFigure}
\end{figure}
\vspace{3mm}

\vspace{1mm}
{\bf Local Operators}.  
Let's now turn our attention to the operators entering eq.\,(\ref{fundamentalmatrixelements}). In a QFT a local operator $\Omega(x,t)$ can be characterised by its matrix elements on the multi-particle states 
\begin{eqnarray}
\tilde F^{\Omega}_{n, m}(\alpha,\beta; x,t) &\,=\,& \langle n, \alpha \mid \Omega(x,t) \mid m, \beta\rangle \,=\, 
\langle n, \alpha \mid e^{-i P x + i H t} \,\Omega(0,0)\, e^{i P x - i H t} \mid  m, \beta\rangle 
\label{FormFactors}\\
& \,=\, & e^{-i (P_\alpha - P_\beta) x + i (E_\alpha - E_{\beta}) t} \, \langle n, \alpha \mid \Omega(0,0)\, \mid m, \beta\rangle 
\,\,\equiv \, e^{-i (P_\alpha - P_\beta) x + i (E_\alpha - E_{\beta}) t}\, F^{\Omega}_{n, m}(\alpha,\beta) \nonumber
\end{eqnarray}
where we have extracted the explicit dependence on $x$ and $t$ of these matrix elements using the space-time translated 
operators $\Omega(x,t) \,=\,e^{i P^\mu x_\mu} \Omega(0,0) e^{-i P^\mu x_\mu}$.  The quantities $F^{\Omega}_{n, m}(\alpha,\beta)$ 
are the so-called Form Factors of the operator $\Omega$, graphically represented as in Figure \ref{FormFactorFigure}: they 
uniquely characterise the various operators. These matrix elements are solutions of a set of functional equations -- known as 
Watson equations \cite{KWSmirnov} -- coming from the unitarity and crossing symmetry of any QFT.  For the two-dimensional integrable 
SUSY theories, the computation of the Form Factors has been discussed in \cite{1998_Mussardo_NPB_532} and, for the purposes of this 
paper, we do not address further this theme, implicitly assuming that an exact computation of the matrix elements of various operators 
can be always performed.   

\vspace{3mm}

Putting now together the various pieces of this discussion, it is easy to see that all the expectation values (\ref{fundamentalmatrixelements}) can be computed once we specify: 
\begin{itemize}
\item the basis $| n,\alpha\rangle$, satisfying the completeness relation (\ref{completeness}); 
\item the amplitudes $K_n(\alpha)$ of the virtual emission of $n$-particle state out of the initial state $| B \rangle$; 
\item the Form Factors $F^{\Omega_i}_{n, m}(\alpha,\beta)$ of the various operators.
\end{itemize} 
The employment of all these quantities provides the {\em spectral representation} of the various correlation functions. For example, with $t_1 \leq t_2$, 
such spectral representation of the two-point functions is given by   
\begin{eqnarray}
&& G_{ab}(x_1,t_1;x_2,t_2 \mid t) \,=\, \langle B(t) \mid \Omega_a(x_1,t_1) \Omega_b(x_2,t_2) \mid B(t) \rangle \nonumber 
\\
& & \,\,\, \,=\,  
\int d\gamma \sum_{r} \langle B(t) \mid \Omega_a(x_1,t_1) \mid r,\gamma \rangle \, \langle r,\gamma \mid \Omega_b(x_2,t_2) \mid B(t) \rangle \\
 & & \,\,\, \,=\,  
\int d\alpha \,d\beta \,d\gamma \,\sum_{n,m,r} K^*_{n}(\alpha) K_m(\beta) 
F^{\Omega_a}_{n, r}(\alpha,\gamma) F^{\Omega_b}_{r, m}(\gamma,\beta)
e^{i  P_\gamma (x_1 -x_2) - i E_\gamma (t_1 - t_2) + i E_\alpha (t_1 -t) - i E_\beta (t_2 - t)} \,\,\,,\nonumber 
\end{eqnarray} 
and, as anticipated, this expression involves the Form Factors $F^{\Omega_i}_{n, r}(\alpha,\gamma)$ of the two operators, the amplitudes $K_m(\alpha)$ of the boundary states together with the energies and momenta $(E_i, p_i)$ of the particles of the spectrum. Obviously an important issue concerns the convergence properties of these series as well as the most efficient way to handle them, but this is somehow a separate issue with respect to the key considerations made here, which were aimed instead to identify the building blocks out of which one can derive in principle the entire out of equilibrium dynamics. 

\section{Questions for SUSY theories out of equilibrium}\label{section3}

In the topic of isolated quantum systems brought out of equilibrium, a decisive question concerns the situation that is reached at $t \rightarrow\infty$, 
in particular to establish whether the system equilibrates or not, and the nature of this equilibrium phase. If the system reaches indeed an equilibrium situation at large time scale, this will be described by a stationary density matrix $\rho_{eq}$ so that 
\be
\lim_{t\rightarrow \infty}  \frac{
\langle B(t) \mid \Omega_a(x_1,t_1)  \ldots \Omega_s(x_n,t_n) \mid B(t) \rangle}{\langle B \mid B \rangle}
\,=\,{\rm Tr} \, \left(\rho_{eq} \Omega_a(x_1,t_1)  \ldots \Omega_s(x_n,t_n)\right) \,\,\,.
\ee
An important issue involves the symmetries present during the time evolution and those which also persist asymptotically in the equilibrium situation.  
In this paper, in particular, we are concerned with the dynamics out of equilibrium of QFT's that are supersymmetric, namely theories that admit an invariance under the exchange of bosonic and fermionic excitations.  In order to express in clear terms what are the interesting aspects and the novelties of this problem, let's briefly summarise here the basics of SUSY, skipping for the time being all technical details which will nevertheless be addressed later. So, let's denote by $Q$ an operator which turns a bosonic state into a fermion states and viceversa
\be 
  Q |\textrm{boson}\rangle \,= \, |\textrm{fermion}\rangle \hspace{8mm} , \hspace{8mm}  Q |\textrm{fermion}\rangle \,=\, |\textrm{boson}\rangle \,\,\,.
  \label{bosonfermioncorr}
\ee
We assume that $Q$ and its hermitian conjugate partner $Q^\dagger$ are symmetries of the systems, i.e. 
they commute with the Hamiltonian $H$ and the momentum $\vec{P}$ 
\be
  [Q, \vec{P}] = [Q, H] = 0 \hspace{6mm} , \hspace{6mm}  [Q^\dagger, \vec{P}] = [Q^\dagger, H] = 0 \,\,\,.
\label{commutationHP}
\ee
Both $Q$ and $Q^\dagger$ are anti-commuting fermionic operators and therefore supersymmetry concerns more with the Lorentz group and the space-time symmetry (associated to the stress-energy tensor and the relative momentum operator $P^\mu$) rather than an internal symmetry. In particular, their anti-commutator is essentially expressed by the Hamiltonian $H$ 
\be 
\{ Q, Q^\dagger\} \simeq H \,\,\,.
\label{fundamentalanticommutator}
\ee
This fact alone is sufficient to explain why SUSY has certain distinguished features with respect to ordinary internal symmetries. Even though they are well known, for the purpose of completeness let's recall and discuss them in what will be called later the {\em zero-temperature situation}, alias in the bulk. 

\vspace{3mm}
{\bf Non-negative energies}. 
The first feature is that all states in a SUSY theory have a non-negative energy: this is a simple consequence of eq. (\ref{fundamentalanticommutator}) since for any state $| \alpha \rangle$ 
we have 
\be 
E_{\alpha}\,=\, \langle \alpha | H | \alpha \rangle \,=\,  \langle \alpha | \{ Q, Q^\dagger\} | \alpha \rangle  \,=\, 
\mid\mid Q\mid \alpha \rangle \mid\mid^2 \, \geq 0 \,\,\,.
\label{nonnegative}
\ee 

\vspace{3mm}
{\bf Multiplets}. 
The second important feature is that, in SUSY theories, particles must be organised into multiplets: since the SUSY generators have half-integer spin under the Lorents transformations, the particles in the multiplets will have different spins but equal mass, since eq.(\ref{commutationHP}) implies that $Q$ and $Q^\dagger$ commute with $P^2$. These multiplets give rise to irreducible representation of the SUSY algebra. 

\vspace{3mm}
{\bf Spontaneously Symmetry Breaking}. 
The third important feature concerns the spontaneous symmetry breaking of a SUSY theory, a condition which only depends upon the energy $E_{GS}$ of its ground state $\mid {\rm GS} \rangle$. This is a {\em if} and {\em only if} condition. Indeed, if SUSY is unbroken, the ground state $\mid {\rm GS}\,\rangle$ of the system must be invariant under SUSY transformations  
\begin{eqnarray}
Q \mid {\rm GS} \, \rangle \,=\,Q^\dagger \mid {\rm GS} \, \rangle\,=\,0,
\end{eqnarray}
and therefore, from  eq.(\ref{nonnegative}), it follows  that the ground state energy $E_{GS}$ mush vanish, $E_{GS} =0$.  
Viceversa, if $E_{GS} \neq 0$, using once again eq.\,(\ref{fundamentalanticommutator}), we see that $Q$ and $Q^\dagger$ 
do not annihilate the ground state $\mid {\rm GS}\,\rangle$ and therefore SUSY must be broken. For this reason, the ground state energy $E_{GS}$ 
is often used as an order parameter to determine whether SUSY is spontaneously broken or not. Moreover, when SUSY is broken, it is well known that this breaking is accompanied by a massless fermionic particle, the so-called Goldstino \cite{Goldstino}, highlighted by a pole singularity at $k^\mu =0$ in the Green function made of fermionic operators.  

\vspace{3mm}
After this short summary of the main features of zero-temperature or bulk SUSY theories, let's now turn back our attention to the quench processes and single out what are the novel questions that emerge in this out of equilibrium situation: we will see that, in a way or another, they are all related to the issue whether SUSY is broken or unbroken after a quantum quench. 

\begin{enumerate}
\item We have already commented that the physical situation which is closer to a quench process is the finite temperature one. Let's see then 
what is the most important point in the finite temperature context: assume that SUSY is exact in the zero-temperature situation, so that in the bulk theory there is a perfect balance between fermions and bosons; once the theory is put at finite temperature, this balance seems however destroyed, since the familiar Fermi or Bose distributions for the occupation number of these particles at energy $E$ are given by 
\be
n_{F,B}(E) \,=\,\frac{1}{e^{\beta E} \pm 1} \,\,\,, 
\label{fbdistributions}
\ee
and they are obviously different. On the basis of this argument, several authors have previously argued that SUSY can never be exactly realised at finite temperature \cite{1978_Das_PRD_12,Girardello}. For analogy, this should also be true after a quantum quench. But is it really so? 
\item Rephrasing the question: assuming that SUSY is unbroken in the zero-temperature case, is the finite density energy $\epsilon_0$ present 
in any quench protocol, see eq.\,(\ref{finitedensity}), responsible for its breaking in the out of equilibrium situation? If so, will the breaking of SUSY be necessarily accompanied by a Goldstino? What will be the nature of this particle? 
\item Even assuming that a quench protocol leads indeed to a breaking of SUSY, will the mass degeneracy between fermions and bosons be necessarily removed?
\item Would it be possible at all to identify a set of amplitudes $K_n(\alpha)$ of the boundary states $|B\,\rangle$ such that the dynamics out of equilibrium will result SUSY invariant?    
\end{enumerate}
Finding answers to these questions will be the subject of the sections that follow. It may be useful, however, to anticipate and underline some of the results in order to guide the reader through the rest of the paper. 

\begin{itemize}
\item 
Concerning question 1, it is important to stress that the Fermi and Bose distributions given in (\ref{fbdistributions}) are essentially free theory concepts. Therefore, the right question to pose is whether there exist {\em interacting} theories where the mode occupations of fermions and bosons -- dynamically and exactly computed -- are still the same at finite temperature. Although it cannot be excluded that this may happen in higher dimensional SUSY theories, this is definitely the case for $d=2$ models and for very good reasons: namely, in $d=2$, the statistical properties of the particles, usually determined by the permutation behaviour of their wave functions, get however mixed up with the interaction. Therefore for this low-dimensional class of theories the usual distinction between boson and fermion assumes another meaning than, say, in ordinary $d=4$ space-time. For this reason, we find interesting to concentrate from now our attention on $d=2$ SUSY theories because for these models there are promising expectations to set 
up a dynamics out of equilibrium which is also supersymmetric.

\item About question 2, as argued in \cite{1984_Matsumoto_PRD_29} and discussed in more detail in Section \ref{FiniteTemperature}, the finite density energy $\epsilon$ induces indeed a breaking of SUSY and a relative singularity at $p^\mu =0$ in the Green functions involving fermionic operators. However, unless SUSY was already broken at zero temperature, the novelty is that this singularity at $p^\mu=0$ has to be interpreted as hydrodynamic singularity rather than a pole due to an intermediate massless fermionic particle, i.e. the Goldstino. So, in the out of equilibrium situations we have the peculiar situation to have a breaking of SUSY but not accompanied by the Goldstino. These considerations have an important consequence, which was not a-priori obvious: the basis of the states that one has to use in the dynamics out of equilibrium is exactly the same as in the zero-temperature situation.  

\item Concerning question 3, we will see that even if SUSY is broken in the quantum quench processes, this does not necessarily imply the removal of degeneracy between bosons and fermions belonging to the same multiplet. So, unless proved differently, we can still assume that a mass degeneracy holds between these two type of particles. 

\item Finally, about question 4, we will see that it is possible to impose certain conditions on the boundary state $| B\,\rangle$ (that obviously 
translate into particular requirements on the $K_n(\alpha)$ amplitudes) in order to ensure a dynamics SUSY invariant.  
 
\end{itemize} 
We will spend the remaining sections to show these results, starting by a more precise presentation on SUSY in $d=2$ and its consequences.

\section{More on d=2 SUSY QFT}\label{moreond=2}
As argued above, the natural place where to look for a perfect balance between fermionic and bosonic distribution also at finite energy density is 
the set of $d=2$ SUSY theories. For this reason here we recall the main formulas relative just for these theories, referring the reader to the more specialised literature for the discussion of the SUSY formalism in all other cases \cite{WZ,Sonhius,Weinberg,Polchinski,mirror,OliveWitten}. 

\vspace{3mm}
\noindent
{\bf $\gamma$ matrices}. Dealing with fermions, the first thing to do is to introduce a proper set of $\gamma$ matrices: in $d=2$ there are only two of such matrices, $\gamma^0$ and $\gamma^1$, and they are both $2\times 2$ dimensional matrices. 
There is also the $\gamma^5$ matrix given by $\gamma^5 = \gamma^0 \gamma^1$. In the following for these matrices we adopt  the 
Majorana representation given by 
\EQ
\gamma^0 \,=\,\sigma_2 \,=\,\left(
\begin{array}{rr}
0 & - i \\
i & 0 
\end{array} 
\right)
\hspace{3mm}
;
\hspace{3mm}
\gamma^1 = i \sigma_1 \,=\,\left(
\begin{array}{rr}
0 &  i \\
i & 0 
\end{array} 
\right)
\hspace{3mm}
,
\hspace{3mm}
\gamma^5 \,=\, \gamma^0 \gamma^1 \,=\, \sigma_3 \,=\,
\left(
\begin{array}{rr}
1 &  0 \\
0 & -1
\end{array} 
\right)
\,\,\,.
\EN
In $d=2$ a generic fermion operator $\psi$ is associated to a two-dimensional spinor $\psi = \left(\begin{array}{l} \psi_+\\\psi_- \end{array}\right)$ and 
we define, as usually, $\bar\psi = \psi^\dagger \gamma^0$. We can introduce the charge-conjugation matrix $C$ satisfying the condition 
\[
(C \gamma^0) \,(\gamma^{\mu})^* \, (C \gamma^0)^{-1} \, =\, - \gamma^\mu \,\,\,.
\] 
It is easy to see that a solution of this equation is given by $C = \gamma^0$: this operator maps the fermion $\psi$ to its conjugate particle $\psi_c$ according to 
\EQ
\psi_c \,=\, (C \gamma^0) \psi^* \,\,\,.
\label{conjugationfermion}
\EN
In this representation a Majorana fermion (which satisfies the neutrality condition $\psi_c \,=\,\psi$) has then both components real 
\EQ
\psi_+^* \,=\, \psi_+ 
\hspace{5mm}
,
\hspace{5mm}
\psi_-^* \,=\, \psi_- 
\,\,\,.
\EN  

\vspace{3mm}
\noindent
{\bf N=1 SUSY Algebra.} 
After all these definitions, let's now introduce the most general $N=1$ SUSY algebra in $d=2$, built up in terms of a single 
Majorana fermionic operator
\be 
Q = \left(\begin{array}{l} Q_+\\Q_- \end{array}\right) \,\,\,,
\ee
satisfying the anti-commutation relation \cite{OliveWitten}
\EQ
\{Q_{\alpha},\bar Q_{\beta}\} \,=\,2 (\gamma_{\lambda})_{\alpha\beta}\,P^{\lambda} + 
2 i (\gamma_5)_{\alpha\beta}\,{\cal Z} \,\,\,,
\label{susyalgebra}
\EN 
where $P^{\lambda}$ is the momentum operator while ${\cal Z}$ is $c$-number called the {\em topological charge}.  Using the explicit 
representation of the $\gamma$-matrices and the Majorana condition for the charge, the anti-commutation relation above translates into the 
conditions
\be
Q_+^2 \,=\, P_+ \hspace{5mm},\hspace{5mm}
Q_-^2 \,=\,P_-\hspace{5mm} , \hspace{5mm}
\{Q_+, Q_-\} \,=\, {\cal Z} \,\,\,. 
\ee
where $P_{\pm} = P^0 \pm P^1$ are the light-cone components of the momentum operator.

\vspace{3mm}
\noindent
{\bf Superfields.} 
Concerning the operator content and the irreducible representations of the SUSY theories, there are of two types: 
the Neveu-Schwartz and the Ramond representations. Here it is useful to discuss only the Neveu-Schwartz representations 
where the bosonic and fermionic fields, together with a real auxiliar field $F(x)$, can be conveniently organised into a real 
superfield  $\Phi(x,\theta)$ that admits the expansion 
\EQ
\Phi(x,\theta) \,=\,\varphi(x) + \bar\theta\,\psi(x) + \frac{1}{2} \bar\theta\,\theta F(x) \,\,\,.
\EN 
The space coordinates $x^{\mu}=(x^0,x^1)$ and the two real Grassmann coordinates $\theta_{\alpha} = (\theta_1,\theta_2)$ 
describe the $N=1$ superspace. A translation in superspace 
\EQ
 x^{\mu} \rightarrow x^{\mu} + i \bar\epsilon \gamma^{\mu} \theta 
\,\,\,\,\,\,\,\,
,
\,\,\,\,\,\,\,\,
\theta_{\alpha} \rightarrow \theta_{\alpha} + \epsilon_{\alpha} 
\label{translationsusy}
\EN 
induces a variation of the superfield given by 
\EQ
\delta \Phi(x,\theta)\,=\,\bar \epsilon_{\alpha} \,{\cal Q}_{\alpha}\,\Phi(x,\theta) \,\,\,,
\label{susytran}
\EN 
with ${\cal Q}_{\alpha} = \partial/\partial \bar\theta_{\alpha} + i (\gamma^{\mu} \theta)_{\alpha}\,
\partial_{\mu}$. The most general action involving a NS superfield and invariant under the supersymmetric transformation 
(\ref{susytran}) can be expressed as  
\EQ
{\cal A} \,=\,\int d^2 x\, d^2 \theta \left[
\frac{1}{4} (\bar D_{\alpha} \Phi) \,D_{\alpha} \Phi + W(\Phi) \right] \,\,\,,
\label{susyaction}
\EN 
where $\int d^2\theta \, \bar \theta \theta = 2$ and the covariant derivative $D_{\alpha}$ 
is given by 
\EQ
D_{\alpha} \equiv \frac{\partial}{\partial \bar\theta_{\alpha}} - 
(i \partial_{\mu} \gamma^{\mu} \theta)_ {\alpha} \,\,\,.
\EN 
$W(\Phi)$ is the so-called superpotential, that we assume to be an analytic function of $\Phi$. 
Integrating on the Grassmann variables, one gets   
\EQ
{\cal A}\,=\,\int d^2 x \left\{
\frac{1}{2} \left[ (\partial_{\mu} \varphi)^2 + i \bar\psi \gamma^{\mu} \partial_{\mu}
\psi + F^2 \right] + F\,W^{'}(\varphi) - \frac{1}{2} W^{''}(\varphi) \bar\psi\psi \right\}\,\,\,,
\EN 
where $W^{'}(\varphi) = dW(\varphi)/d\varphi$, etc.  Finally, eliminating the auxiliary field $F$ from 
its algebraic equation of motion, i.e. substituting $F \rightarrow - W^{'}(\varphi)$ in the above 
expression (the so-called {\em on-shell SUSY}), and rescaling for convenience the fermion field as $\psi \rightarrow \sqrt{2} \psi$, 
it yields the general form of the lagrangian density for a supersymmetric theory given by
\EQ
{\cal L} \,=\,
\frac{1}{2} \left[ (\partial_{\mu} \varphi)^2 - [W^{'}(\varphi)]^2\right]  
+ i \bar\psi \gamma^{\mu} \partial_{\mu}
\psi  - W^{''}(\varphi) \bar \psi \psi \,\,\,.
\label{finalsusy}
\EN
Associated to the transformation (\ref{translationsusy}) there is the conserved supercurrent 
\EQ
J^{\mu}_{\alpha}(x) \,=\,(\partial_{\nu} \varphi) (\gamma^{\nu} \gamma^{\mu} \psi)_{\alpha} - i F 
(\gamma^{\mu} \psi)_{\alpha} \,\,\,, 
\label{supercurrent}
\EN 
and the associated supercharges 
\EQ
Q_{\alpha} \,=\,\int dx^1\,J_{\alpha}^0 \,\,\,.
\label{supercharge}
\EN 
Let's also define the stress-energy tensor 
\EQ
{\cal T}^{\mu\nu}(x) \,=\,i\,\bar\psi \,\gamma^{\mu} \partial^{\nu} \psi + 
\partial^{\mu} \varphi \,\partial^{\nu} \varphi - \frac{1}{2} g^{\mu\nu} \left[
(\partial_{\alpha} \varphi)^2 - F^2\right] 
\,\,\,,
\label{stressenergy}
\EN 
and the topological current 
\EQ
\xi^{\mu}(x) \,=\,-\epsilon^{\mu\nu} F\,\partial_{\nu} \varphi \,=\,\epsilon^{\mu\nu} \partial_{\nu} W(\varphi) 
\,\,\,.  
\EN 
In terms of these quantities we have 
\be 
P^{\lambda} \,=\, \int {\cal T}^{0 \lambda}(x) \,dx^1
\ee
and 
\EQ
{\cal Z}_{ab} \,=\,\int \xi^0(x)\, dx^1 \,=\,\left[W(\varphi)\right]^{+\infty}_{-\infty} \,\equiv\,
W(\varphi_b) - W(\varphi_a)  \,\,\,, 
\label{topologicalcharge}
\EN 
where $\varphi_a$ and $\varphi_b$ are vacuum (constant) configurations of the theory. So ${\cal Z}_{ab} \neq 0$ only if the superpotential $W(\varphi)$ has a degenerate set of vacua, connected to each other by solitons. Using now all these definitions, it is easy to see that the supercharges (\ref{supercharge}) close indeed the supersymmetry algebra (\ref{susyalgebra}). 

\vspace{3mm}
\noindent
{\bf Integrable SUSY Models}. Particularly interesting SUSY theories for our considerations are those which have, in addition to the conserved 
supercharges, also an infinite number of bosonic charges ${\cal I}_n$ that commute among themselves and with the supercharges. If this is the case, we are in presence of  integrable theories that can be fully characterised in terms of their elastic and factorisable $S$-matrix, as briefly discussed in the next section. An example of such a kind of theories is given by the SUSY Sine-Gordon model \citep{SSG}, whose superpotential $W(\Phi)$  
\EQ
W(\Phi) \,=\, \frac{m}{\lambda^2} \,\cos(\lambda \Phi) \,\,\,
\EN 
ends up to the on-shell Lagrangian 
\EQ
{\cal L} \,=\,\frac{1}{2} (\partial_\mu\varphi)^2 + \frac{m^2}{2 \lambda^2} \,\sin^2\lambda\varphi + i \bar\psi\gamma^\mu\partial_\mu\psi -  
m \bar\psi \psi \cos\lambda \varphi \,\,\,.
\label{SUSYSG}
\EN
This theory has several degenerate ground states at $\varphi = n \pi/\lambda$ and therefore the anti-commutation of its supercharges involves a non-zero values of the topological charge $\mathcal{Z}_{ab}$.

\vspace{3mm}
\noindent
{\bf Models with hidden SUSY}. Interestingly enough, in $d=2$ there may exist theories which present SUSY but only as hidden symmetry. In particular, 
even though these theories do not admit a lagrangian formulation based on the superfield formalism presented above, their dynamics is 
nevertheless ruled by an underlying set of fermionic operators $Q_{\pm}$ closing the SUSY algebra (\ref{susyalgebra}). Examples of this kind of theories are:
\begin{itemize} 
\item the {\em ordinary} Sine-Gordon model (not to be confused with the SUSY SG above), for a very special value of its coupling constant, i.e. $g^2=16\pi/3$; at this value the spectrum of the model consists of soliton and anti-solitons alone which, we will see in the next Section, set up an irreducible representation of both $N=2$ and $N=1$ SUSY algebra \cite{1998_Mussardo_NPB_532,DenisAndre}. 
 \item the Tricritical Ising Model perturbed by its vacancy density operator along its first-order phase transition line \cite{ZamTIM}: the theory has three degenerate vacua connected by kink excitations, without any additional bound states. These kinks give rise to an irreducible  representation of the SUSY algebra and their scattering amplitudes are purely elastic and factorizable. 
\end{itemize} 
In view of their potential experimental realisation, these two models will be among our main subjects of interest in the sequel.

\section{SUSY at finite energy density}\label{FiniteTemperature}

In this section, essentially following  the approach proposed in \cite{1984_Matsumoto_PRD_29}, we are going to show that:
\begin{enumerate}
\item[{\bf a}] SUSY is broken at finite energy density; 
\item[{\bf b}] the correlation functions involving fermion operators have a singularity at $p^\mu =0$; 
\item[{\bf c}] this singularity at $p^\mu$, however, is not due to the presence of a massless fermionic particle (the familiar Goldstino) but has to be interpreted instead as some collective excitation of the media, the so-called {\em phonino}, i.e. a supersymmetric sound wave \cite{2003_Kratzer_AP_308}. 
\end{enumerate}

For what we are concerned in this paper, the crucial result is that the phonino is actually not a particle excitation and therefore we do not need to insert into a resolution of the identity in terms of the states of a basis. Said differently, the existence of the phonino is a purely emergent phenomena and moreover its existence is independent of the space-time dimensionality of the theory. This result is of outmost importance for understanding SUSY out of equilibrium, since it settles the important question relative to the basis that has to be used to describe the Hilbert space in the out of equilibrium dynamics of the theory: this basis is made of exactly the same particle states entering the theory in the bulk and, if SUSY was unbroken in the bulk, there is no need to take into account extra excitations due to the Goldstino. 

Let's sketch the main points which are necessary to establish these results, referring the reader to the literature also quoted below for a more detailed discussion. Our discussion starts from the familiar argument to establish the existence of the Goldstino when SUSY is spontaneously broken \cite{Goldstino}. Consider the Ward-Takahashi identity involving the supercurrent $J^{\mu}(x)$ 
\be 
\partial_{\mu} \langle \bar\epsilon J^{\mu}(x) \phi_1(x_1) \ldots \phi_n(x_n) \rangle \,=\,
\sum_{k=1}^n \langle \phi_1(x_1) \ldots \phi_{k-1}(x_{k-1}) \delta\phi_k(x_k) \ldots \phi_n(x_n) \rangle \,\delta^{d}(x-x_k) \,\,\,, 
\ee
where 
\be
\delta \phi(x) \,=\,\epsilon \{Q,\phi(x)\} \,\,\,.
\ee 
The spontaneously breakdown of SUSY is realised when the ground state is not annihilated by the supercharges and therefore 
some fermionic operator $\psi$ transforms inhomogeneously  
\be
\langle \{Q,\psi(x) \} \rangle \neq 0 \,\,\,. 
\ee 
In this case we have 
\be
\partial_{\mu} \langle J^{\mu}(x) \psi(y) \rangle\,=\,\delta^d(x-y) \,\langle \{Q,\psi(y)\} \rangle \,\,\,,  
\label{gammamu}
\ee
which, taking the Fourier transform
\be 
\Gamma^{\mu}(p) \,=\,\int d^d x e^{i p (x-y)} \, \langle J^{\mu}(x) \psi(y) \rangle \,\,\,,
\ee
can be expressed as 
\be 
p_\mu \, \Gamma^{\mu}(p) \,\neq 0 \,\,\,. 
\label{polegold}
\ee
This says that the correlation function $\Gamma^{\mu}(p)$ has a pole at $p^\mu = 0$. For the bulk theory, where we have Lorentz invariance,  
eq.\,(\ref{polegold}) implies that there must be a pole for all light-like momenta $p^2= 0$, i.e. there must exist a massless Goldstone particle. However, given the non-covariant structure of the quench process or the finite temperature (there is a preferred direction, alias the time axis, in both situations), one cannot draw this conclusion in these cases. The nature of the pole present in $\Gamma^{\mu}(p)$ needs a different explanation.  

For the finite temperature case, such an explanation can be found employing the thermo-field dynamics formalism \cite{1984_Matsumoto_PRD_29,Dasbook} and this will serve our scopes too, given the close relation between the finite-temperature and quantum quench cases. In the thermo-field dynamics one would like to express the thermal averages in terms of expectation values on the ``thermal state" $\mid 0,\beta \rangle$, as much as it is done in the zero-temperature case, where all computations reduce to compute expectation values on the vacuum state. So, for a generic observable ${\mathcal O}$ we pose 
\be 
\langle {\mathcal O} \rangle_\beta \,=\,Z^{-1}(\beta) \sum_{n} \langle n | {\mathcal O} | n \rangle\,e^{-\beta E_n} \equiv 
\langle 0, \beta | {\mathcal O} | 0,\beta \rangle \,\,\,.
\label{thermalvacuum}
\ee 
The problem is then to identity the thermal vacuum. It is easy to see that it cannot be simply express in terms of the states $| n\rangle$. 
Indeed, if we pose 
\be
| 0, \beta \rangle \,=\,\sum_{n} c_n \,| n \rangle \,\,\,,
\ee
we have 
\be
\langle 0, \beta | {\mathcal O} | 0,\beta \rangle \,=\, 
\sum_{n,m} c^*_m \, c_n \,\langle m | {\mathcal O} | n\rangle \,\,\,, 
\ee
and comparing now with eq.\,(\ref{thermalvacuum}), we see that it must hold the identity
\be 
c^*_m \,c_n  \,=\,Z^{-1}(\beta) \, e^{-\beta E_n} \,\delta_{m,n} \,\,\,,
\ee
which is impossible, since $c_n$'s are ordinary numbers.  Hence, as long as we restrict to the original Hilbert space we cannot define a thermal vacuum so that it holds the identity (\ref{thermalvacuum}). 

The way out of this nuisance is to double the Hilbert space by introducing a fictitious ancilla copy of the original Hilbert space, denoted by tilde, and write the thermal state as 
\be 
| 0, \beta \rangle \,=\,\sum_n c_n \,|n \rangle \otimes | \tilde  n \rangle \,\,\,.
\ee
The role of the ancilla states is simply to enforce the orthogonality condition: in fact, using this expression we have 
\begin{eqnarray} 
\langle 0, \beta | {\mathcal O} | 0,\beta \rangle & \,=\,&  
\sum_{n,m} c^*_m \, c_n \,\langle m | {\mathcal O} | n\rangle \,\langle \tilde m | \tilde n \rangle \nonumber \\
&\,=\,& \sum_n c^*_n \, c_n \,\langle n | {\mathcal O} | n\rangle 
\end{eqnarray}
and now there is of course no problem in posing 
\be 
\mid c_n \mid^2 \,=\, Z^{-1}(\beta) \, e^{-\beta E_n} \,\,\,.
\ee
What discussed above are the basic footsteps of the so-called thermo-field dynamics. In the case of field excitations, as shown in \cite{1984_Matsumoto_PRD_29}, the states $| \tilde n\rangle$ have to be taken with both energy and momentum opposite to the usual particles therefore the spectral density $\sigma(k^0,\vec{k})$ associated to $\Gamma^{\mu}(k)$ assumes the form 
\be
\sigma(p^0,\vec{p}) \propto \sum_{m,n} \langle 0,\beta | J^\mu(0) | n,\tilde m\rangle \langle n,\tilde m |\psi(0)| 0,\beta\rangle 
\delta(p^0 - E_n + E_m) \,\delta^{d-1}(\vec{p} - \vec{p}_n + \vec{p}_m) \,\,\,,
\ee
therefore a zero-energy and zero-momentum singularity is caused by a collective set of thermal pairs of excitations. A low-energy divergence of response functions due to collective excitations is a signature of sound waves. As shown in \cite{2003_Kratzer_AP_308}, this is also the case for the spontaneous breaking of SUSY at finite temperatures. Therefore, ultimately, the singularity usually attributed to the goldstino has to be instead  interpreted as due to the supersymmetric sound waves.

\section{Elastic SUSY  S-matrix}\label{ElasticSUSYSMatrix}

Quantum integrability in $d=2$ implies that the scattering processes are elastic and factorizable. In this case, all the scattering information of the theory is encoded in the two-particle S-matrix. The general $N=1$ SUSY S-matrices of this kind were extensively discussed by Schoutens in  \cite{SUSYschoutens}. For simplicity, let's initially consider theories whose spectrum consists of only one species of boson and fermion, both of mass $m$, without additional internal indices. A one-particle state will be denoted by $| b(\beta) \rangle$ or $ | f(\beta) \rangle$, depending whether boson or fermion, or generically as $ \vert A(\theta)\rangle$. The parameter $\theta$ is the particle's rapidity, entering the dispersion relations $E=m\cosh\theta,\,p=m\sinh\theta$. The SUSY charges act on these states as 

\EQ
\begin{array}{l}
{Q}_+ \mid b(\beta) \rangle = \omega \,\,\,\,\sqrt{m} e^{\beta/2} 
\mid f(\beta)\rangle \,\,\,;\\ 
{Q}_+ \mid f(\beta) \rangle = \omega^{-1} \sqrt{m} e^{\beta/2} 
\mid b(\beta)\rangle\,\,\,;
\end{array}
\,\,\, \,\,\, 
\begin{array}{l}
 { Q}_- \mid b(\beta) \rangle = \rho \,\,\,\,\sqrt{m} e^{-\beta/2} 
\mid f(\beta)\rangle \,\,\,;\\ 
{ Q}_- \mid f(\beta) \rangle = \rho^{-1} \sqrt{m} e^{-\beta/2} 
\mid b(\beta)\rangle \,\,\,,
\end{array}
\label{Qoneparticle}
\EN 
i.e. in terms of two matrices 
\EQ
{\cal Q}_+ = \left(
\begin{array}{cc} 
0 & \omega \\
\omega^{-1} & 0 \\
\end{array}
\right)
\,\,\,\,\, , \,\,\,\,\, 
{\cal Q}_- = \left(
\begin{array}{cc} 
0 & \rho \\
\rho^{-1} & 0 \\
\end{array}
\right) 
\,\,\, ,
\label{Q+-}
\EN 
satisfying ${\cal Q}_+^2 = {\cal Q}_-^2 =1$ and $\{{\cal Q}_+, {\cal Q}_-\}=0$ (i.e. there is no topological charge),  where 
\be
\omega = - i \rho \,=\, e^{i\pi/4}\,\,\,.
\ee 
The action of ${Q}_+$ and ${ Q}_-$ on a multi-particle states must take into account the fermionic nature of these operators 
and therefore involves brading relations
\begin{eqnarray}
&& { Q}_+  
\mid A_1(\beta_1) A_2(\beta_2) \ldots A_n(\beta_n) \rangle = 
\sqrt{m} \sum_{k=1}^n e^{\beta_k/2} 
\label{braidingQ} 
\\ 
&& \,\,\,
\mid (Q_F A_1(\beta_1)) (Q_F A_2(\beta_2)) \ldots 
(Q_F A_{k-1}(\beta_{k-1}) (Q_+ A_k(\beta_k))
A_{k+1}(\beta_{k+1}) \ldots A_n(\beta_n) \rangle  
\nonumber
\end{eqnarray}
and 
\begin{eqnarray}
&& { Q}_- 
\mid A_1(\beta_1) A_2(\beta_2) \ldots A_n(\beta_n) \rangle = 
\sqrt{m} \sum_{k=1}^n e^{-\beta_k/2} 
\label{braidingQbar} \\ 
&& \,\,\,
\mid (Q_F A_1(\beta_1)) (Q_F A_2(\beta_2)) \ldots 
(Q_F A_{k-1}(\beta_{k-1}) (Q_- A_k(\beta_k))
A_{k+1}(\beta_{k+1}) \ldots A_n(\beta_n) \rangle  
\nonumber
\end{eqnarray}
where $Q_F$ is the fermion parity operator, which on the basis $\mid b\rangle$ and $\mid f\rangle$ is represented by the 
diagonal matrix 
\EQ
{\cal Q}_F = \left(\begin{array}{cc}
1 & 0 \\
0 & -1 
\end{array}
\right) \,\,\, . 
\label{Q_L}
\EN  
Particularly important for what follows is the representation of the two super-charges  on the two-particle states $\mid b(\beta_1) b(\beta_2)\rangle$,  
$\mid f(\beta_1) f(\beta_2)\rangle$, $\mid f(\beta_1) b(\beta_2)\rangle$, $\mid b(\beta_1) f(\beta_2)\rangle$. The first two states belong to the 
$F=1$ sector (even number of fermionic particles) whereas the remaining two states to the $F=-1$ sector (odd number of fermionic particles). By choosing for convenience $\beta_1 =\beta/2$ and $\beta_2 =-\beta/2$, 
the operator ${Q}_+$ will be represented by the matrix 
\EQ
{\cal Q}_+(\beta) = 
\left(\begin{array}{cccc}
0 & 0 & \omega x & \omega x^{-1} \\
0 & 0 & -\omega^{-1} x^{-1} & \omega^{-1} x \\
\omega^{-1} x & -\omega x^{-1} & 0 & 0 \\
\omega^{-1} x^{-1} & \omega x & 0 & 0 
\end{array}
\right) 
\,\,\, ,
\label{q2particle}
\EN 
where $x\equiv e^{\beta/4}$. For ${Q}_-$ we have analogously 
\EQ
{\cal Q}_-(\beta) = 
\left(\begin{array}{cccc}
0 & 0 & \rho x^{-1} & \rho x \\
0 & 0 & -\rho^{-1} x & \rho^{-1} x^{-1} \\
\rho^{-1} x^{-1} & -\rho x & 0 & 0 \\
\rho^{-1} x & \rho x^{-1} & 0 & 0 
\end{array}
\right) \,\,\,.
\label{qbar2particle}
\EN 
In the following we will also need the representation matrix of the operator ${Q}_+ Q_-$ on the above 
two particle states, given by  
\EQ
({\cal Q}_+ {\cal Q}_-)(\beta) = 2  
\left(\begin{array}{cccc}
\frac{\omega}{\rho} & -\omega \rho \sinh\frac{\beta}{2}& 0 & 0 \\
\frac{1}{\omega\rho} \sinh\frac{\beta}{2} & -\frac{\omega}{\rho} & 0 & 0\\
0 & 0 & 0 & -\frac{\omega}{\rho} \cosh\frac{\beta}{2} \\
0 & 0 & -\frac{\omega}{\rho} \cosh\frac{\beta}{2} & 0 
\end{array}
\right) \,\,\,.
\label{qqbarparticle}
\EN 
The two-particle S-matrix is defined by
\begin{eqnarray}
S^{l k}_{ij}(\theta)\vert A_l(\theta_2), A_k(\theta_1)\rangle_{\rm out}=\vert A_{i}(\theta_1), A_j(\theta_2)\rangle_{\rm in}\,\,\,,
\label{definitionSmatrix}
\end{eqnarray}
where $\theta=\theta_1-\theta_2$. For SUSY invariant theories, in addition of the conditions coming from unitarity, crossing symmetry, analyticity and Yang-Baxter equations, these amplitudes must be also invariant under the action of the supercharges, which leads to the equation 
\begin{eqnarray}
&& {\cal Q}_\pm(\theta)\,S_{ij}^{lk}(\theta)\,=\,S_{ij}^{lk}(\theta)\,{\cal Q}_\pm(-\theta),\,\,,\label{sysuss1}\\
& & ({\cal Q}_+ {\cal Q}_-)(\theta)\, S_{ij}^{lk}(\theta)\,=\,S_{ij}^{lk}(\theta)\,({\cal Q}_+ {\cal Q}_-)(-\theta) \label{sysuss2}\,\,\,.
\end{eqnarray}
Interesting examples of $S$-matrices which satisfy these conditions, even though their excitations are not obviously of fermionic or bosonic type, 
are those of the Sine-Gordon model at a particular value of its coupling constant and the Tricritical Ising Model, briefly recalled hereafter. The former and the latter model give rise to systems of zero and non-zero SUSY topological charge ${\mathcal Z}$ respectively. 

\vspace{3mm}
\noindent 
{\bf Sine-Gordon model}. 
The simplest possible SUSY $S$-matrix, solution of the eqs.(\ref{sysuss1}) and (\ref{sysuss2}), and satisfying unitarity and crossing symmetry equation 
is a $4 \times 4$ matrix that, written in the basis $|bb\rangle$, $|ff\rangle$, $|fb\rangle$, $|bf\rangle$ basis, reads \cite{1998_Mussardo_NPB_532}
 \be 
  S(\theta) \,=\, R(\theta) \begin{pmatrix}
    -1/\cosh(\theta/2) & i\tanh(\theta/2) & 0 & 0 \\
    i\tanh(\theta/2) & -1/\cosh(\theta/2) & 0 & 0 \\
    0 & 0 & -1 & 0 \\
    0 & 0 & 0 & -1
    \end{pmatrix}
    \label{SGSUSYS}
\ee
with 
\ea {
  R(\theta) = \exp\left[\frac{i}{2} \int_0^{\infty} \frac{dt}{t} \frac{\sin\frac{\theta t}{\pi}}{\cosh^2 \frac{t}{2}}\right]\,\,\,.
}
As a matter of fact, this $S$-matrix is identical to the ordinary Sine-Gordon $S$-matrix for the soliton/antisolitons at the special value of the 
coupling $g^2=16\pi/3$. Here $g$ is the interaction parameter appearing in the Sine-Gordon Lagrangian
\ea {
  \mathcal{L}_{SG} \,=\, \frac{1}{2}\left(\partial_{\mu} \phi\right)^2 + \frac{m^2}{g^2}\cos(g\phi)\,\,\,. 
}
The Lagrangian of the sine-Gordon model does not explicitly exhibit any sign of SUSY, as it is written in terms of only one bosonic field. However, as we are going to recall, in this model the SUSY operators transforms the solitons and antisolitons into each other,  i.e. SUSY relates different topologically charged, or \textquotedblleft soliton creating \textquotedblright fields. These soliton creating operators were discussed in \cite{lukyanovzamolodchikov}, where their form factors were also computed. They are defined as
\begin{eqnarray}
\mathcal{O}_a^n(x)=\lim_{\epsilon\to +0} \exp\left\{-\frac{n}{4\beta}\int_{-\infty}^x \partial_y \phi(x,y) dx\right\} \exp\left\{i a \phi(x+\epsilon, y)\right\},
\end{eqnarray}
where $n$ is an integer denoting the topological charge of the operator. In the case $n=0$, these reduce to the well-known vertex operators
\begin{eqnarray}
\mathcal{O}_a^0(x)=\exp\left\{i a \phi(x,y)\right\}.
\end{eqnarray}
These soliton creating fields provide an explanation for the emergence of SUSY at the particular value of the coupling. First of all, for any value of the coupling, it was shown in \cite{DenisAndre} that the Sine-Gordon has semilocal conserved charges, that generate the affine quantum group $U_q(\widehat{sl(2)})$, with 
\begin{eqnarray}
q=e^{i 8\pi^2/\beta^2}.
\end{eqnarray}
Secondly, these conserved charges can be written in terms of the soliton creating operators as \cite{lukyanovzamolodchikov}
\begin{eqnarray}
&&G_{\pm}=\frac{1}{N_Q} \int_{-\infty}^\infty \left(\mathcal{O}^{\pm2}_{\pm(2\beta/\sqrt{8\pi})^{-1}}(x)+\pi \xi m \mathcal{O}^{\pm 2}_{\pm \nu}(x)\right) dx,\\
&& \bar{G}_{\pm}=\frac{1}{N_Q} \int_{-\infty}^\infty \left(\mathcal{O}^{\pm 2}_{\mp (2\beta/\sqrt{8\pi})^{-1}}(x)+\pi \xi m \mathcal{O}^{\pm 2}_{\mp \nu}(x)\right) dx,
\end{eqnarray}
where $\xi$ is the usual renormalised coupling constant of the Sine-Gordon model
\begin{eqnarray}
\xi=\frac{\beta^2}{8\pi\left(1-\frac{\beta^2}{8\pi}\right)},
\end{eqnarray}
while $\nu = (2\beta/\sqrt{8\pi})^{-1} - \beta/\sqrt{8\pi}$ and $N_Q$ is a normalization constant. Thirdly, these semilocal conserved charges satisfy the generalized commutation conditions
\begin{eqnarray}
G_{-} \bar{G}_{+}-q^2 \bar{G}_{+}G_{-}=\frac{1-q^{2H}}{1-q^{-2}},\,\,\,\,\,\,G_{+}\bar{G}_{-}-q^2\bar{G}_{-}G_{+}=\frac{1-q^{-2H}}{1-q^{-2}},\label{qgeneralcommutator}
\end{eqnarray}
where $H$ is the usual topological charge of the Sine-Gordon model  
\begin{eqnarray}
H=\frac{\beta}{16\pi^2}\int_{-\infty}^\infty \partial_x \phi(x,y) dx.
\end{eqnarray}
Now it is easy to see what is so special about the SUSY value of the coupling, $\beta^2=16\pi/3$. For this value, in fact, the conditions (\ref{qgeneralcommutator}) simply become anti-commutation conditions 
\begin{eqnarray}
G_{-}\bar{G}_{+}+\bar{G}_{+}G_{-}=0,\,\,\,\,\,\,G_{+}\bar{G}_{-}+\bar{G}_{-}G_{+}=0.
\end{eqnarray}
As discussed in \cite{lukyanovzamolodchikov}, at this value of the coupling these conserved charges have spin $1/2$, and their action on one-soliton (antisoliton) states is given by 
\begin{eqnarray}
&&G_{\pm}\vert A_{\pm}(\theta)\rangle = 0 \,\,\,\,\,\,\,\,\,\,\,\,\, \,\,\,\,\,\,\,\,\,\,\,\,\,\,,\,\,\,\,\bar{G}_{\pm}\vert A_{\pm}(\theta)\rangle = 0,\nonumber\\
&&G_{\mp}\vert A_{\pm}(\theta)\rangle=e^{\frac{\theta}{2}}\vert A_{\mp}(\theta)\rangle\,\,\,,\,\,\,\bar{G}_{\mp}\vert A_{\pm}(\theta)\rangle=e^{-\frac{\theta}{2}}\vert A_{\mp}(\theta)\rangle,\label{qonstates}
\end{eqnarray}
where $\vert A_+(\theta)\rangle$ and $\vert A_-(\theta)\rangle$ are one soliton and one antisoliton states, respectively.
It can be shown that the fermionic charges $G_{\pm},\, \bar{G}_{\pm}$ generate even a larger extended $N=2$ SUSY. 
The $N=1$ SUSY sub-algebra which interests us in this paper is constructed in terms of the semilocal charges as 
\begin{eqnarray} \label{W_def}
W_+=G_++G_-\,\,\,\,\,\,W_-=-i\bar{G}_+ + i\bar{G}_-,
\end{eqnarray}
In the following sections we will construct a boundary state related to the $N=1$ SUSY of the model, rather than its extended 
original $N=2$ SUSY.

Coming back now to the expression of the $S$-matrix (\ref{SGSUSYS}), in the Sine-Gordon model the basis involving solitons and anti-solitons is given by $|S\bar{S}\rangle$, $|\bar{S} S\rangle$, $|SS\rangle$, $|\bar{S}\bar{S}\rangle$. The two basis, one of bosons and fermions and the other of solitons and antisolitons, can be identified with each other in four possible ways: we can identify, for instance,  $|bb\rangle$ with $|S\bar{S}\rangle$ but also with $|\bar{S}S\rangle$; this choice forces a unique identification of $|ff\rangle$ with either $|\bar{S}S\rangle$ or $|S\bar{S}\rangle$. In the same way, and independently, we have two ways of identifying $|bf\rangle$ and $|fb\rangle$ with $|SS\rangle$ and $|bf\rangle$. As we shall see later this ambiguity leads to the same physical situation.

Notice that even though the solitons and anti-solitons are topological excitations of the Sine-Gordon model, from the SUSY point of view they are considered instead to have ${\mathcal Z}= 0$, i.e. zero SUSY topological charge. An example of non-zero SUSY topological charge is given by the 
Tricritical Ising Model.

\begin{figure}[t]
\psfig{figure=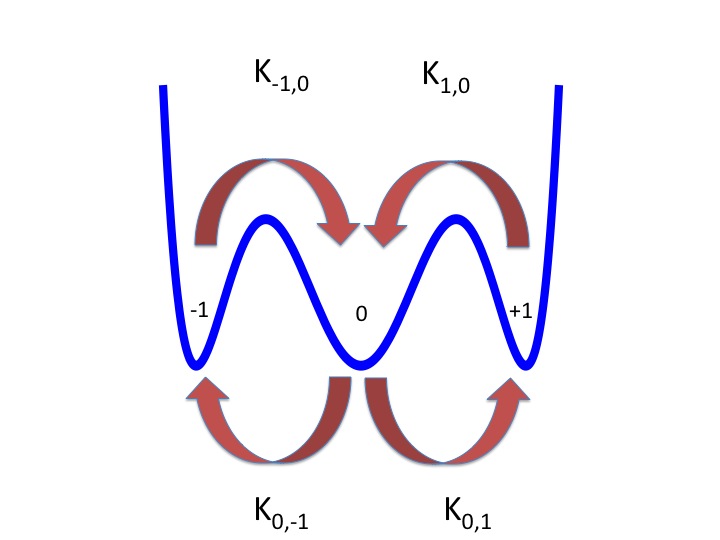,height=5cm,width=9cm}
\caption{{\em Effective potential for the order parameter $\sigma(x)$ of the Tricritical Ising Model along the first order phase transition line. 
There are three degenerate vacua connected by the kink excitations $\mid K_{ab}(\theta)\rangle$.}}
\label{TIMvacua3}
\end{figure}

\vspace{3mm}
\noindent 
{\bf Tricritical Ising Model}. An interesting statistical field theory model which shows a SUSY invariance in the scattering 
amplitudes of its excitation is the Tricritical Ising Model (TIM) once perturbed along its vacancy density operator \cite{ZamTIM,ZamRSOS}. This model 
can be also formulated in such a way to exhibit its SUSY invariance also at its critical point \cite{Qiu,MSS}. This permits to organise its operator content (usually classified only in terms of the irreducible representations of the Virasoro algebra) in terms of a superfield 
\be \label{TIM_superfield}
\Phi(x,\theta) \,=\,\epsilon(x) + \bar\theta \psi(x) + \frac{1}{2} \bar\theta \theta t(x) \,\,\,,
\ee
where, in addition to the fermion $\psi(x)$, there are the operator $\epsilon(x)$ -- the energy density operator -- and the field $t(x)$ which describes  
the vacancy density. All these operators belong to the spin $Z_2$ even sector of the model. In this model there are two additional $Z_2$ odd fields, $\sigma(x)$ and $\sigma'(x)$ which play the role of magnetization and sub-leading magnetization operators respectively: they also give rise to other irreducible representations of the SUSY but in the so-called Ramond sector \cite{Qiu,MSS}.  Perturbed by the vacancy operator $t(x)$, the effective SUSY off-critical action is given by the Landau-Ginzburg action 
\begin{eqnarray}
 {\cal A} &\,=\, & \int d^2 x\, d^2 \theta \left[
\frac{1}{4} (\bar D_{\alpha} \Phi) \,D_{\alpha} \Phi + \frac{1}{3!} \Phi^3 + \lambda \Phi \right] \nonumber \\
&\,=\,& \int d^2 x \left(
\frac{1}{2} \left[ (\partial_{\mu} \epsilon)^2  
+ i \bar\psi \gamma^{\mu} \partial_{\mu}
\psi \right]   - (\epsilon + \lambda)^2 -  \epsilon \bar \psi \psi\right) \,\,\,.
\label{susyactionTIM}
\end{eqnarray}
When $\lambda >0$ there is a massless flow from TIM to the Ising model and since $\lambda > 0$ gives rise to a spontaneously breaking of SUSY,  the corresponding Goldstino in this case is just the familiar Majorana fermion of the Ising model \cite{Martinec}. On the other hand, when $\lambda < 0$, SUSY is exact but there is a degeneracy of the vacuum, which results to be doubly degenerate if described in terms of the $\epsilon$ variable, but 
triple degenerate if expressed instead in terms of the order parameter $\sigma(x)$, as it comes from the operator identity 
$\epsilon = :\sigma^2 :$ \cite{ZamTIM}.  The potential for the field $\sigma$ is plotted in Figure \ref{TIMvacua3}: the farthest vacua are denoted by $\pm 1$ while the central one denoted by $0$; in the Figure there are also drawn the kink excitations $| K_{ab}(\theta)$ which connect the various neighboring vacua. 
In this case we have four one-particle states: $| K_{-,0} \rangle$, $| K_{+,0}\rangle$, $ | K_{0,-}\rangle$ and $| K_{0,+}\rangle$, and the three matrices
${\cal Q}_+, {\cal Q}_-$ and ${\cal Q}_F$ on these states take the form 
\be \label{TIM_matrix_Q}
{\cal Q}_+ \,=\, 
\left(\begin{array}{rrrr}
0 & i & 0 & 0 \\
-i & 0 & 0 & 0 \\
0 & 0 & 1 & 0 \\
0 & 0 & 0 & -1 
\end{array}
\right) 
\hspace{3mm}
, 
\hspace{3mm}
{\cal Q}_- \,=\, 
\left(\begin{array}{rrrr}
0 & i & 0 & 0 \\
-i & 0 & 0 & 0 \\
0 & 0 & -1 & 0 \\
0 & 0 & 0 & 1 
\end{array}
\right) 
\hspace{3mm}
, 
\hspace{3mm}
{\cal Q}_F \,=\, 
\left(\begin{array}{rrrr}
0 & 1 & 0 & 0 \\
1 & 0 & 0 & 0 \\
0 & 0 & 0 & 1 \\
0 & 0 & 1 & 0 
\end{array}
\right) \,\,\,.
\ee
Notice that the topological charge ${\mathcal Z}$ on the one-particle states this time is different from zero  
\be
{\mathcal Z} \,=\,\{{\cal Q}_+,{\cal Q}_-\} \,=\, 
\left(
\begin{array}{rrrr}
2 & 0 & 0 & 0 \\
0 & 2 & 0 & 0 \\
0 & 0 & -2 & 0 \\
0 & 0 & 0 & -2
\end{array}
\right) 
\ee
and, indeed, on the kink $| K_{ab}\rangle$ connecting the vacua $a$ and $b$ the topological charge ${\mathcal Z}$ takes the value $2 (a^2 - b^2)$.   
The two-body elastic $S$-matrix for the kink excitations is defined by ($\theta_{12} \equiv \theta_1-\theta_2$)
\be
\mid K_{ac}(\theta_1) K_{cb}(\theta_2) \rangle \,=\, S_{ab}^{cd} (\theta_{12}) \, \mid K_{ad}(\theta_2) K_{db}(\theta_2) \rangle \,\,\,,
\ee
and the non--zero amplitudes --fixed in terms of SUSY invariance, continuity of the vacuum indices in the kink states, crossing and unitarity equations -- are given by \cite{ZamTIM} 

\vspace{3mm}
\hspace{25mm}\begin{picture}(190,40)
\thicklines
\put(60,0){\line(1,1){30}}
\put(60,30){\line(1,-1){30}}
\put(60,15){\makebox(0,0){$0$}}
\put(90,15){\makebox(0,0){$0$}}
\put(75,0){\makebox(0,0){$\pm $}}
\put(75,30){\makebox(0,0){$\pm $}}
\put(100,15){\makebox(0,0)[l]{$\displaystyle{
\,\,\,\, =\,\,\,S_{00}^{\pm  \pm }(\theta)\,=\,
\sigma(\theta) \,e^{-i \rho\theta}\, \cosh\frac{\theta}{4} }
$}}
\put(390,15){\makebox(0,0)[l]}
\end{picture}

\vspace{3mm}

\hspace{25mm}\begin{picture}(190,40)
\thicklines
\put(60,0){\line(1,1){30}}
\put(60,30){\line(1,-1){30}}
\put(60,15){\makebox(0,0){$0$}}
\put(90,15){\makebox(0,0){$0$}}
\put(75,0){\makebox(0,0){$\pm $}}
\put(75,30){\makebox(0,0){$\mp $}}
\put(100,15){\makebox(0,0)[l]{$\displaystyle{
\,\,\,\, =\,\,\, S_{00}^{\pm  \mp }(\theta)\,=\,
- i \sigma(\theta) \,e^{-i \rho\theta}\,  \sinh\frac{\theta}{4}} 
\label{STIMSUSY}$}}
\put(390,15){\makebox(0,0)[l]}
\end{picture}

\vspace{3mm}

\hspace{25mm}\begin{picture}(190,40)
\thicklines
\put(60,0){\line(1,1){30}}
\put(60,30){\line(1,-1){30}}
\put(60,15){\makebox(0,0){$\pm $}}
\put(90,15){\makebox(0,0){$\pm $}}
\put(75,0){\makebox(0,0){$ 0 $}}
\put(75,30){\makebox(0,0){$0$}}
\put(100,15){\makebox(0,0)[l]{$\displaystyle{
\,\,\,\, =\,\,\, S_{\pm \pm }^{0 0 }(\theta)\,=\, \sigma(\theta) \,e^{i \rho\theta}\, \left(\cosh\frac{\theta}{4} + i  \sinh\frac{\theta}{4}\right)} 
$}}
\put(390,15){\makebox(0,0)[l]}
\end{picture}

\vspace{3mm}

\hspace{25mm}\begin{picture}(190,40)
\thicklines
\put(60,0){\line(1,1){30}}
\put(60,30){\line(1,-1){30}}
\put(60,15){\makebox(0,0){$\pm $}}
\put(90,15){\makebox(0,0){$\mp $}}
\put(75,0){\makebox(0,0){$ 0 $}}
\put(75,30){\makebox(0,0){$0$}}
\put(100,15){\makebox(0,0)[l]{$\displaystyle{
\,\,\,\, =\,\,\,  S_{\mp \pm }^{0 0 }(\theta)\,=\, \sigma(\theta) \,e^{i \rho\theta}\, \left(\cosh\frac{\theta}{4} - i  \sinh\frac{\theta}{4}\right)} 
$}}
\put(390,15){\makebox(0,0)[l]}
\end{picture}

\vspace{3mm}
\noindent 
where $\rho \,=\, (1/2\pi) \log 2$ while the function $\sigma(\theta)$ which implements the unitarity condition 
reads 
\be
\sigma(\theta) \,=\, \left(\cosh\frac{\theta}{2}\right)^{-1/2}\, 
\exp\left[\frac{i}{4} \,\int_0^\infty \frac{dt}{t} \frac{\sin\frac{t \theta}{\pi}}{\cosh^2\frac{t}{2}}\right]
\,\,\,.
\label{sigmafactorTIM}
\ee
Notice that, denoting by $z_i$ ($i=1,2$) the SUSY topological charges of each excitation, the two-particle states entering the non-zero scattering 
amplitudes are all and only those with vanishing total SUSY topological charge ${\mathcal Z} = z_1 + z_2$ \cite{SUSYschoutens}.  
As a rule of thumb, the role of bosonic and fermionic excitations of this theory are played by those who diagolize both $Q_F$ and ${\mathcal Z}$
\be
\begin{array}{lll}
| b_+\rangle \simeq \left(| K_{-0}\rangle + | K_{+0}\rangle\right) &\hspace{2mm}, \hspace{2mm} & |f_+\rangle  \simeq 
\left(| K_{-0}\rangle  - | K_{+0}\rangle\right) \\
& & \\
| b_- \rangle \simeq \left(| K_{0-}\rangle + | K_{0+} \rangle\right) & \hspace{2mm} , \hspace{2mm} & | f_+\rangle  \simeq 
\left(| K_{0-}\rangle  - |  K_{0+}\rangle\right)  
\end{array}
\ee
even though, as explained in \cite{SUSYschoutens}, these assignments could not be taken literally in view of the kink nature of the individual excitation.

\section{Boson and Fermion Occupation Number in d=2 theories}\label{Bose&Fermion in d=2}

In light of all we have learned in the previous sections, let's briefly go back to an important issue mentioned in Section \ref{section3}. In that Section,  
we recalled that one of the main argument against the possibility to implement an exact SUSY theory at finite temperature was the difference  
between the thermal Fermi or Bose distributions for the occupation number of these particles at energy $E$, given by 
\be
\rho_{F,B}(E) \,=\,\frac{1}{e^{\beta E} \pm 1} \,\,\,. 
\label{fbdistributions22}
\ee
In Section \ref{section3}, however, we also argued that these familiar Bose/Fermi distributions are expressions referring to free theories while the actual computation of the 
occupation number distributions must take into account the interaction among the particles. Therefore it cannot be excluded that in  
{\em interacting } SUSY theories, there could be a perfect balance between boson and fermion occupation numbers. This statement 
may be difficult to prove for generic $d$-dimensional theories, the problem is indeed open, but here we would like to point out that 
this is indeed what may happen quite naturally in $d=2$ theories:  for integrable theories, the way to prove such a statement passes by 
the derivation of the entire thermodynamics using the exact two-body $S$-matrix \cite{ZamTBA}.  Without entering into many details, for which 
we refer the reader to the literature on Thermodynamics Bethe Ansatz \cite{ZamRSOS,ZamTBA} and, in more details, to the papers that implemented 
such an approach in the SUSY case \cite{TBASUSY}, let's simply mention that  in $d=2$ theories the statistical properties are dictated 
by the value of the $S$-matrix at zero rapidity, $S(0)$: remarkably, all known $S$ matrices which correspond to interacting theories 
have $S(0) = -1$ (fermionic type) and therefore all particles behave essentially as fermions. Indeed, looking at the definition (\ref{definitionSmatrix}), it is clear that the $S$-matrix plays the role of braiding operator as well.

Put the theory on a finite volume, of length $L$ along the space $x$-direction and $R=1/T$ along the time $t$-direction, the Hilbert space 
of the integrable models split into subspaces of assigned number $N$ of particles, whose allowed rapidities $\{\theta_j\}_{j=1}^N$ obey the 
quantization condition implemented by the Bethe equations (suppressing the S-matrix indices for simplicity) \cite{TBASUSY} 
\ea { \label{Bethe}
  e^{i p_j L} \prod_{j\neq k} S(\theta_j-\theta_k)_{jk} = -1,\quad\quad j=1,\dots, N \,\,\,.
}
Here $p_j=m_j\sinh(\theta_j)$ is the relativistic momentum. Because the S-matrix is of the fermionic type all the rapidities are different and there is no solution of eq.~\eqref{Bethe} if two or more rapidities coincide. The state of a thermal equilibrium is determined by an extremum of the free energy $F = E - T\mathcal{S}$, where $E$ is the energy of the particle configurations while $\mathcal{S}$ their entropy. In the thermodynamic limit $L\rightarrow \infty$ with $N/L$ finite, this extremum condition leads --- dynamically --- to the determination of the occupation number 
distributions $\rho_a(\theta)$ of the particle $a$, expressed as 
\ea {
  \rho_a^{(p)}(\theta) \equiv \frac{1}{1+\exp(-\epsilon_a(\theta))} \,\,\,,
}
through a set of non-linear integral equations for the functions $\epsilon_a(\theta)$ (called pseudo-energies) which generalize -- at finite temperature -- the relativistic form $m_a \cosh(\theta)$ of the energy of the particle $a$.   

Hence, if the pseudo-energies $\epsilon_a(\theta)$ are identical for all the particles in the supermultiplets, then there will be a perfect equivalence between bosonic and fermionic excitations even at non-zero temperature, contrary to the naive perturbative argument. Notice that this equivalence is pretty obvious in the case of the Sine-Gordon model, where the roles of boson and fermion are essentially played by the soliton and anti-soliton of the theory: in absence of a chemical potential that weights differently these two excitations, their occupation number $\rho_a(\beta)$ has to be identical by charge conjugation. The equivalence between bosonic and fermionic occupation number is also respected for the Tricritical Ising Model \cite{ZamRSOS}. 

\section{Supersymmetric Boundary States for Quench Dynamics}\label{SUSYBOUNDARY}

Having established in the previous sections that, at least in $d=2$, a finite energy density does not preclude a SUSY dynamics out of equilibrium 
(since the SUSY breaking firstly does not spoil the bulk basis and secondly does not preclude a perfect balance of bosonic and fermionic populations), we now move on to address another building blocks of the dynamics out of equilibrium, alias the boundary states $| B \rangle$ and the associated particle amplitudes $K_n(\alpha)$. In particular, here we would like to identify certain boundary states $| B \rangle$ which ensure a SUSY dynamics out of the equilibrium. It is worth underline that the conditions discussed below are only sufficient and not necessary, meaning that it is not excluded that 
there could be a dynamical restoration of SUSY even though this symmetry may be explicitly broken by the form of the boundary state. This issue about SUSY restoration through the time evolution will be further discussed in our final Section \ref{conclusions}: here, instead we focus our attention on the identification of a particularly simple class of SUSY boundary states. These will be the states associated to one of the two following conditions 
\begin{eqnarray}
(Q_+ \pm i Q_-)\vert B\rangle=0\,\,\, .\label{susyconstraint}
\end{eqnarray}
Notice that, if imposed at $t=0$, these equations also hold at any later time 
\begin{eqnarray}
(Q_+ \pm i Q_-)\vert B(t) \rangle=0\,\,\, ,\label{susyconstraint22}
\end{eqnarray}
since $|B(t) = e^{-i H t} |B\rangle$ and the Hamiltonian $H$ commutes with the SUSY charges. 

Let's justify these equations. Their main motivation comes from the general analysis of two-dimensional integrable QFT's with a boundary \cite{GZ} that, as well known, admit two equivalent formulations: (i) in the first, the boundary is placed at $t=0$ (i.e. is a {\em time boundary}), the Hilbert space is the same as in the bulk and therefore the boundary plays the role of initial condition for the system, associated to a particular state $| B \rangle$; (ii) in the second, the boundary is instead placed at $x=0$ (is a {\em space boundary}), the theory is defined only on the half space $x > 0$, the Hilbert space of states is not the one of the bulk, since particle states with negative momenta are not independent but are related, through  the reflection scattering amplitudes, to those with positive momenta. The former formulation, where we have a time boundary, is of course the one relevant for the quench dynamics. Let's then initially consider an integrable system with the boundary in time, at $t=0$: in this case the boundary state $| B \rangle$ satisfies the conditions \cite{GZ}
\begin{eqnarray}
(P_+^{(s)} -P_-^{(s)})\vert B\rangle=0\,\,\,,\label{integrableboundary}
\end{eqnarray}
for all the integer-spin $s$ associated to the conserved charges in the bulk. In SUSY theories there are additional half-integer-spin conserved 
charges $Q_{\pm}$ and it is easy to see that the conditions (\ref{susyconstraint}) can be interpreted as the \textquotedblleft square root \textquotedblright of the $s=1$ condition
\begin{eqnarray}
(P_+^{(1)} - P_-^{(1)})\vert B\rangle \,=\, (Q_+ \pm i Q_-)(Q_+ \pm i Q_-)\vert B\rangle\,=\,0\,\,\,.
\end{eqnarray}
This is of course not a derivation of the conditions (\ref{susyconstraint}) but rather a step forward the understanding of their meaning. To better justify them, let's turn now the attention to the case where the boundary is instead in space, at $x=0$. In this case the space translation invariance is obviously broken.  Hence, in the presence of a SUSY invariance in the bulk, one can expect to have at most only \textquotedblleft $1/2$ \textquotedblright of the original supersymmetric invariance. With the assumption that the two components of the SUSY charges are on the same footing, being these 
components made of real Majorana fermions, the simplest consistent boundary conditions that can be put at $x=0$ are 
\be 
(Q_+ \pm Q_-)\mid_{x=0} \,=\, 0 \,\,\,. 
\ee
These conditions were used, for instance, to find SUSY boundary reflection matrices, as discussed in \cite{nepomechie,nepomechieTIM}.
We can now perform a space-time rotation, interchanging the role of $x$ with $t$ and transforming the two components of the SUSY 
charges in the new frame: as a result of this crossing transformation, one gets our original conditions (\ref{susyconstraint}) for the boundary state 
$| B \rangle$. It must be said that the entire discussion is completely analogous to the one for the Majorana fermion fields in the Ising model with boundary, a situation analysed in \cite{GZ}.   

For an integrable QFT to which are imposed boundary conditions which are also integrable, the structure of the boundary state is further constrained by the infinite number of bosonic conservation laws (\ref{integrableboundary}), among which the space component of the momentum and all higher odd powers thereof. In light of these conservation laws, the boundary state $| B \rangle$ must be made of pairs of particles of equal mass with opposite rapidities, which can be therefore expressed as
\begin{eqnarray}
&\vert B\rangle\,=\,\sum_{N=0}^\infty \int_{0<\theta_1<\theta_2<\dots<\theta_N} d\theta_1 d\theta_2\dots d\theta_N \, K_{2N}^{a_n \dots a_1, b_1 \dots b_N}(\theta_1,\theta_2,\dots,\theta_N)\nonumber\\
&\nonumber\\
&\vert A_{a_N}(-\theta_N) \ldots A_{a_1}(-\theta_1)\,A_{b_1}(\theta_1)\ldots A_{b_N}(\theta_N)\rangle \,\,\,.\label{cooperpairs}
\end{eqnarray}
Moreover, using once again the integrability, this expression can be further elaborated and the boundary state can be neatly written in terms of the two-particle component as \cite{GZ} 
\begin{eqnarray}
\vert B\rangle\,=\,\exp\left(\int_{-\infty}^\infty d\theta K(\theta)\right)\vert 0\rangle \,\,\,, 
\label{exponentialboundarystate}
\end{eqnarray}
with 
\begin{eqnarray}
K(\theta)\,=\,\frac{1}{2}K^{ij}(\theta)A^\dag_i(-\theta)A^\dag_j(\theta)\,=\,\frac{1}{2}K^{ij}(-\theta)A^\dag_i(\theta)A^\dag_j(-\theta)\,\,\,,
\end{eqnarray}
where $A^\dag_{i}(\theta)$ are the particle creation operators. The aim of this Section is to show that the SUSY condition (\ref{susyconstraint}) imposes further constraints on the structure of the boundary state (\ref{cooperpairs}), in particular on the two-particle amplitudes $K^{ij}(\theta)$. In the rest of this section we consider supersymmetric theories that have only one species of boson and one fermion.

Since the boundary state of an integrable theory is composed of pairs of particles with opposing rapidities, we need to analyse how the supercharges act on these two-particle states: 
\begin{eqnarray}
Q_+\vert A_{i}(\theta) A_{j}(-\theta)\rangle &\,=\,& \sqrt{m}\left(x\,\mathcal{Q}_+\otimes\mathbf{1}+x^{-1} \mathcal{Q}_F\otimes\mathcal{Q}_+\right)\vert A_{i}(\theta)A_{j}(-\theta)\rangle,\\
Q_- \vert A_{i}(\theta) A_{j}(-\theta)\rangle &\,=\, & \sqrt{m}\left(x^{-1}\,\mathcal{Q}_-\otimes\mathbf{1}+x \mathcal{Q}_F\otimes\mathcal{Q}_-\right)\vert A_{i}(\theta)A_{j}(-\theta)\rangle,
\end{eqnarray}
where $x\equiv e^{\beta/4}$, while the matrices $\mathcal{Q}_\pm$ and $\mathcal{Q}_{L}$ are defined in (\ref{Q+-}) and (\ref{Q_L}) respectively. Combining these equations to express the constraints (\ref{susyconstraint}) we get 
\begin{eqnarray}
\left(Q_+ \pm i Q_- \right)\, && \vert A_i(\theta)A_j(-\theta)\rangle \, = \, \sqrt{m}\left[\left(x\mathcal{Q}_+ \pm i x^{-1} \mathcal{Q}_- \right) 
\otimes\mathbf{1} + \mathcal{Q}_F \otimes \left(x^{-1} \mathcal{Q}_+ \pm i x \mathcal{Q}_-\right) \right]\vert A_i(\theta)A_j(-\theta)\rangle\nonumber\\
&&\,\,\,\,\, = \sqrt{m}\left[\left(\begin{array}{cc} 0 & (x \mp x^{-1})\omega \\ (x \pm x^{-1})\omega^{-1} & 0\end{array} \right) \otimes
\left(\begin{array}{cc}1&0\\0&1\end{array}\right) + \right. \nonumber \\
&& \,\,\,\,\,\,\,\,\,\,\,\,\,\,\,\,\,\,\,\,\,\,\,\,\,\,\,\,\,\,\,\,\,\,\,\,\,\left.
+\,\left(\begin{array}{cc}1&0\\0&-1\end{array}\right)\otimes\left(\begin{array}{cc} 0 & (x^{-1}\mp x)\omega \\ (x^{-1}\pm x)\omega^{-1} & 0 \end{array}\right)\right]\vert A_i(\theta)A_j(-\theta)\rangle\,\,\,.
\end{eqnarray}
On the other hand, since the general two-particle state can be expressed as
\begin{eqnarray}
\vert A_{i}(\theta) A_j(-\theta)\rangle \,&=&\, \alpha \,\vert b(\theta) b(-\theta) \rangle + \beta\, \vert f(\theta) f(-\theta)\rangle + 
\gamma \,\vert b(\theta) f(-\theta)\rangle  + \delta \,\vert f(\theta) b(-\theta)\rangle \\
&\equiv&\, \alpha\, \begin{pmatrix} 1 \\ 0\end{pmatrix}\otimes \begin{pmatrix} 1 \\ 0\end{pmatrix} + \beta\, \begin{pmatrix} 0 \\ 1\end{pmatrix}\otimes \begin{pmatrix} 0 \\ 1\end{pmatrix} + 
\gamma \,\begin{pmatrix} 1 \\ 0\end{pmatrix}\otimes \begin{pmatrix} 0 \\ 1\end{pmatrix}  + \delta \,\begin{pmatrix} 0 \\ 1\end{pmatrix}\otimes \begin{pmatrix} 1 \\ 0\end{pmatrix}
\,\,\,, \nonumber
\end{eqnarray}
let's find out the conditions on the coefficients $\alpha ,\beta ,\gamma ,\delta $, such that (\ref{susyconstraint}) is satisfied. The analysis 
involves separately the sectors with different fermion number.  

\vspace{3mm}
\noindent 
{\bf  Sector with fermion number $F=1$.}  For the state belonging to the fermion number $F=1$ we have 
\begin{eqnarray}
&\alpha\,(Q_+ \pm i Q_-) \vert b(\theta) b(-\theta)\rangle = \alpha \omega^{-1}\sqrt{m}\left[\left(\begin{array}{c}0\\x\pm x^{-1}\end{array}\right)\otimes\left(\begin{array}{c}1\\0\end{array}\right)+\left(\begin{array}{c}1\\0\end{array}\right)\otimes\left(\begin{array}{c}0\\x^{-1}\pm x\end{array}\right)\right]\nonumber\\
&\nonumber\\
&=\alpha\omega^{-1}\sqrt{m}\left[(x\pm x^{-1})\left(\begin{array}{c}0\\1\end{array}\right)\otimes\left(\begin{array}{c}1\\0\end{array}\right)+(x^{-1}\pm x)\left(\begin{array}{c}1\\0\end{array}\right)\otimes\left(\begin{array}{c}0\\1\end{array}\right)\right],\nonumber\\
&\nonumber\\
&\nonumber\\
&\beta \,(Q_+ \pm i Q_-) \vert f(\theta) f(-\theta)\rangle=\beta\omega\,\sqrt{m}\left[\left(\begin{array}{c}x\mp x^{-1}\\0\end{array}\right)\otimes\left(\begin{array}{c}0\\1\end{array}\right)+\left(\begin{array}{c}0\\-1\end{array}\right)\otimes\left(\begin{array}{c}x^{-1}\mp x\\0\end{array}\right)\right]\nonumber\\
&\nonumber\\
&=\beta \omega\sqrt{m}\left[(x\mp x^{-1})\left(\begin{array}{c}1\\0\end{array}\right)\otimes\left(\begin{array}{c}0\\1\end{array}\right)-(x^{-1}\mp x)\left(\begin{array}{c}0\\1\end{array}\right)\otimes\left(\begin{array}{c}1\\0\end{array}\right)\right]\,\,\,.\nonumber
\end{eqnarray} 
Therefore the contributions coming from the boson-boson and fermion-fermion pairs cancel each other out if it holds the conditions 
\begin{eqnarray}
\alpha \omega^{-1}\left(x \pm x^{-1}\right) \,=\, \beta\omega \left(x^{-1} \mp x\right),\,\,\,\,\,\,\,\,\,\,\,\,\, 
\alpha \omega^{-1}\left(x^{-1}\pm x\right)\,=\, -\beta\omega \left(x\mp x^{-1}\right). \label{abconditions}
\end{eqnarray}
For the state identified by the constraint $(Q_+ + i Q_-) | B \rangle = 0$, both equations become
\begin{eqnarray}
\frac{\alpha}{\beta}\,=\,-i \, \frac{x-x^{-1}}{x+x^{-1}}\,=\, -i \tanh\frac{\theta}{2}.
\end{eqnarray}
while for the state identified by the constraint $(Q_+ - i Q_-) | B \rangle =0 $, the conditions (\ref{abconditions}) become
\begin{eqnarray}
\frac{\alpha}{\beta}\,=\, i\,\frac{x+x^{-1}}{x-x^{-1}}\,=\,  i \coth\frac{\theta}{2}.
\end{eqnarray}

\vspace{3mm}
\noindent
{\bf Sector with fermion number $F=-1$.} For the states belonging to the fermion number $F =1$ we have:
\begin{eqnarray}
&\gamma \,(Q_+ \pm i Q_-) \vert b(\theta) f(-\theta)\rangle = 
\gamma \,\sqrt{m}\left[\omega^{-1}(x\pm x^{-1})\left(\begin{array}{c}0\\1\end{array}\right)\otimes\left(\begin{array}{c}0\\1\end{array}\right)+\omega(x^{-1}\mp x)\left(\begin{array}{c}1\\0\end{array}\right)\otimes\left(\begin{array}{c}1\\0\end{array}\right)\right] \nonumber\\
&\nonumber\\
&\delta \, (Q_+ \pm i Q_-) \vert f(\theta) b(-\theta)\rangle = 
\delta \,\sqrt{m}\left[\omega(x\mp x^{-1})\left(\begin{array}{c}1\\0\end{array}\right)\otimes\left(\begin{array}{c}1\\0\end{array}\right)-\omega^{-1}(x^{-1}\pm x)\left(\begin{array}{c}0\\1\end{array}\right)\otimes\left(\begin{array}{c}0\\1\end{array}\right)\right].\nonumber
\end{eqnarray}
Again, we see that these contributions cancel each other if the following conditions are satisfied:
\begin{eqnarray}
\gamma \left(x \pm x^{-1}\right)\,=\, \delta \left(x^{-1}\pm x\right),\,\,\,\,
\gamma \left(x^{-1} \mp x\right) \,=\, - \delta \left(x\mp x^{-1}\right), 
\end{eqnarray}
whose solution is
\begin{eqnarray}
\gamma \,=\, \pm \delta.
\end{eqnarray}

\vspace{3mm}
\noindent
{\bf Supersymmetric Boundary State}. 
In conclusion, we show that in the $F=1$ sector the SUSY conditions (\ref{susyconstraint}) on the boundary state can only fix the relative ratio of the amplitudes $K_{bb}$ and $K_{ff}$, while in the sector $F= -1$ only the ratio of the amplitudes $K_{bf}$ and $K_{fb}$. Therefore the most general boundary state solution of the SUSY constraints (\ref{susyconstraint}) can be written as
\begin{eqnarray}
\vert B\rangle=\exp&& \left\{ \int_{-\infty}^\infty d\theta \,K_e(-\theta)\left[\left(-i \tanh\frac{\theta}{2}\right)^{\pm 1} A^\dag_b(\theta)A^\dag_b(-\theta) + 
A^\dag_f(\theta)A^\dag_f(-\theta)\right]\right.
\nonumber\\
&& \nonumber\\
&& +\left. \int_{-\infty}^\infty d\theta \,K_o(-\theta)\left[\pm A^\dag_b(\theta)A^\dag_f(\theta) + A^\dag_f(\theta)A^\dag_b(-\theta)\right]\right\}\vert 0\rangle\,\,\,\,,\label{susyboundarystate}
\end{eqnarray}
where the even/odd fermion number amplitudes $K_{e,o}(\theta)$ however are not further constrained by the SUSY conditions (\ref{susyconstraint}). States as those in (\ref{susyboundarystate}) are invariant only under half of the supersymmetry transformations that we denote by $\delta_{\epsilon}$
\ea {
  \delta_{\epsilon_{\pm}}|B\rangle \,=\, e^{i\epsilon_{\pm}(Q_+ \pm i Q_-)}|B\rangle \,=\, |B\rangle\,\,\,.
}
Let's discuss an example of such a boundary state which is provided by the Sine-Gordon model. 

\section{Supersymmetric Boundary State in the Sine-Gordon model}\label{SGBoundary}
In Section \ref{ElasticSUSYSMatrix} we have seen that the exact $S$-matrix of the ordinary Sine-Gordon model at the value 
of the coupling $g^2=16\pi/3$ provides a non-trivial example of SUSY $S$-matrix, with a certain ambiguity of identifying bosons and fermions with solitons and antisolitons. Let us resolve this ambiguity and choose the following mapping
\ea { \label{SineGordon_mapping}
  |bb\rangle = |S\bar{S}\rangle, \quad |ff\rangle = |\bar{S}S\rangle, \quad |bf\rangle = |SS\rangle, \quad |fb\rangle = |\bar{S}\bar{S}\rangle\,\,\,.
}
In the genuine Sine-Gordon model the general boundary state is made of pairs of solitons and anti-solitons
\ea {
 | B \rangle \,=\, \exp\left(\int_{-\infty}^{\infty} d\theta K(\theta) \right) |0\rangle, && K(\theta) = \frac{1}{2}K^{ij}(\theta) A_i^{\dagger}(-\theta) A_j^{\dagger}(\theta), \label{kSG}
}
with indices $i,j$ running over $s$ (soliton) and $\bar{s}$ (antisoliton). The amplitudes $K^{ij}(\theta)$ are now related to the amplitudes of SUSY pairs of particles (made of bosons and fermions) through the assignment~\eqref{SineGordon_mapping}. Therefore imposing the SUSY conditions \eqref{susyconstraint} on the boundary state in the boson/fermion basis, or equivalently imposing
\ea { \label{sineGordon_initial}
  \left(W_+ \pm i W_- \right)|B\rangle = 0
}
with $W_{\pm}$ defined in~\eqref{W_def} in the soliton/antisoliton basis, we get the constrains on the amplitudes
\ea {
  K^{s\bar{s}}(\theta) &= \left(-i\tanh \frac{\theta}{2}\right)^{\pm} K^{\bar{s}s}(\theta),\\
  K^{ss}(\theta) &= \pm K^{\bar{s}\bar{s}}(\theta).
  \label{conditionsss}
}
To unveil the meaning of these conditions, it is worth comparing the expression of these amplitudes with those known from the 
original analysis of the boundary Sine-Gordon model done by Ghoshal and Zamolodchikov \cite{GZ}. In this paper, the authors 
initially consider the boundary in space placed at $x=0$ and compute the reflection amplitudes for a soliton (antisoliton) to scatter off the 
boundary 
\ea { \label{spatial_boundary_amplitudes_SG}
  P_{\pm}(\theta) = \cos(\xi \mp i\lambda \theta) R(-i \theta),\\
  T_{\pm}(\theta) = \frac{k}{2} \sin(-2i\lambda\theta) R(-i \theta).
}
The overall function $R(\theta)$ can be fixed using the boundary unitarity and cross-unitarity equations. The $P_{\pm}(\theta)$ amplitudes describe the diagonal scattering of a soliton (antisoliton) off a boundary into a soliton (antisoliton), while the $T_{\pm}(\theta)$ amplitudes describe the 
scattering channel where the soliton bounces back turning into an anti-soliton and vice-versa. Therefore, if non zero, the amplitudes $T_\pm(\theta)$ signal a violation of charge conjugation symmetry. The parameter $\lambda$ is related to the coupling constant $\beta$: at the SUSY point 
of the model, $g^2=16\pi/3$, it is worth to know that $\lambda=1/2$. The parameters $\xi$ and $k$ entering the reflection amplitudes depend on the boundary term present in the action, and this dependence is, in general, implicit. There are, however, two-limiting cases which are particularly important: these correspond to the free and fixed  boundary conditions. For the free boundary conditions, the amplitudes are pairwise equal ($P_+ = P_-$ and $T_+ = T_-$) while the parameter $\xi$ assumes the value $\xi=0$; for the fixed boundary conditions, instead, the amplitudes $T_{\pm}(\theta)$ vanish ($k=0$) while $P_\pm(\theta)$ continue to depend upon the parameter $\xi$, that is related to the value that the field $\varphi(x)$ of the Sine-Gordon model assumes at the boundary $x=0$: so, for instance, $\xi =0$ corresponds to $\varphi(x=0) = 0$. 

\begin{figure}[t]
\psfig{figure=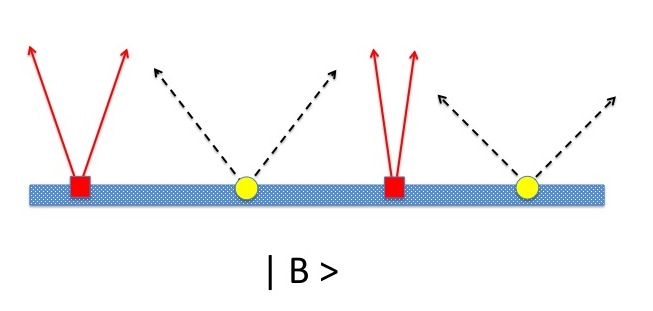,height=6cm,width=12cm}
\caption{{\em SUSY Boundary State in the Sine-Gordon model at $g^2=16\pi/3$, made of virtual emission amplitudes of boson-boson pairs (black-dashed lines) and fermion-fermion pairs (red lines), whose ratio is fixed by eq.\,(\ref{SUSYRATIO}).}}
\label{SGboundaryfigure}
\end{figure}

Swapping now from a boundary placed in space to a boundary placed in time, it is necessary to make a crossing transformation, with  
the incoming particle that becomes its antiparticle: hence, the amplitudes $P_{\pm}(\theta)$ and $T_{\pm}(\theta)$ are in correspondence 
with the amplitudes $K^{ij}(\theta)$ introduced in eq.(\ref{kSG}) as  
\ea {
  P_+\left(i\frac{\pi}{2}-\theta\right) \,=\, K^{s\bar{s}}(\theta), && P_-\left(i\frac{\pi}{2}-\theta\right) \,=\, K^{\bar{s}s}(\theta),\\
  T_+\left(i\frac{\pi}{2}-\theta\right) \,=\, K^{\bar{s}\bar{s}}(\theta), && T_-\left(i\frac{\pi}{2}-\theta\right)\, = \,K^{ss}(\theta).
}
Comparing now with the SUSY conditions (\ref{conditionsss}), we see that we can satisfy these equations for the amplitudes 
\be 
  \frac{P_+\left(i\frac{\pi}{2}-\theta\right)}{P_-\left(i\frac{\pi}{2}-\theta\right)} = \frac{K^{s\bar{s}}(\theta)}{K^{\bar{s}s}(\theta)} = 
  \left(-i\tanh\frac{\theta}{2}\right)^\pm,  
\label{SUSYRATIO}
\ee
by posing $\xi = \pm \pi/4$. So, summarising: while the SUSY conditions does not have realisation in terms of the free boundary condition, they  
can be instead implemented by a particular fine tuning of the fixed boundary conditions for the field, i.e. $\xi = \pm \pi/4$ relative to the two equations 
$(Q_+ \pm iQ_-)|B\rangle = 0$. 

Finally let us comment on the ambiguity of identifying bosons and fermions with solitons and antisolitons. We have seen that this mapping can be achieved in four possible ways but, in view of the results got for the boundary state, the choice of mapping of $|bf\rangle$ into $|SS\rangle$ or $|\bar{S}\bar{S}\rangle$ is irrelevant because the corresponding amplitudes are all zero. We are then left with two possible ways of mapping $|bb\rangle$ into either $|S\bar{S}\rangle$ (as we have assumed above) or into $|\bar{S}S\rangle$. Choosing the second way, this will simply exchange the role of $P_+(\theta)$ with $P_-(\theta) $ and therefore the equations $(W_+  \mp  i W_-)|B\rangle = 0$ correspond to $(Q_+  \pm i Q_-)| B\rangle = 0$, whose 
solutions correspond again to $\xi=\pm\pi/4$. In other words we simply exchange the boundary condition in terms of the $W$'s operators, leaving the amplitudes and $\xi$ invariant.

In conclusion, in the Sine-Gordon model at $g^2=16\pi/3$ for each choice of the sign in equation
\ea {
  \left(Q_+ \pm i Q_- \right)|B\rangle = 0
}
there is one SUSY boundary state $|B\rangle$ that respect the charge-conjugation of the model,
made of a soliton/anti-soliton pairs, alias of an equal mixture of 
boson-boson and fermion-fermion pairs, as shown in Figure \ref{SGboundaryfigure}.

\section{SUSY Boundary State in the Tricritical Ising Model}\label{TIMBoundary}

In this Section we will determine the possible supersymmetric boundary states of the Tricritical Ising Model (TIM). The analysis of this model 
and the identification of such boundary states do not follow from the previous section, since the Tricritical Ising Model has more than one kind of boson and fermion pairs, and a nonzero topological charge $Z$. As shown in Section \ref{ElasticSUSYSMatrix}, the one-particle spectrum consist of four kinks that interpolate between the three adjacent vacua, labelled as $-1,0,+1$. 

Assuming that the boundary state $| B \rangle$ of this model follows the exponential form (\ref{exponentialboundarystate}) discussed in Section 
\ref{SUSYBOUNDARY}, let's see what conditions we get for the particle pair amplitudes from the SUSY equations (\ref{susyconstraint}). 
It is convenient to study first the case where the boundary state has no topological charge, ${\cal Z} \vert B\rangle =0$, and consider later the case where the boundary state has a non-zero topological charge. 

\vspace{3mm}
\noindent
{\bf Boundary State with Zero Topological Charge}. In this case, for a generic two-kink state $\vert A_{i,j}(\theta) A_{j,k}(-\theta)\rangle$, the SUSY condition (\ref{susyconstraint}) is expressed as 
\begin{eqnarray}
&\left(Q_+ \pm  i Q_-\right)\vert A_{i,j}(\theta) A_{j,k}(-\theta)\rangle\nonumber\\
&= \sqrt{m}\left[
\left(\begin{array}{cccc}
0&ix\mp x^{-1}&0&0\\
-ix\pm x^{-1}&0&0&0\\
0&0&x\mp ix^{-1}&0\\
0&0&0&-x\pm ix^{-1}\end{array}\right)\otimes
\left(\begin{array}{cccc}
1&0&0&0\\
0&1&0&0\\
0&0&1&0\\
0&0&0&1\end{array}\right)\right.\nonumber\\
&+\left.\left(\begin{array}{cccc}
0&1&0&0\\
1&0&0&0\\
0&0&0&1\\
0&0&1&0\end{array}\right)\otimes
\left(\begin{array}{cccc}
0&ix^{-1}\mp x&0&0\\
-ix^{-1}\pm x&0&0&0\\
0&0&x^{-1}\mp ix&0\\
0&0&0&-x^{-1}\pm ix\end{array}\right)\right]\nonumber\\
&\times\vert A_{i,j}(\theta) A_{j,k}(-\theta)\rangle=0\,\,\,,
\label{SUSYTIMcondition}
\end{eqnarray}
using the matrix expressions \eqref{TIM_matrix_Q} for $\mathcal{Q}_+$, $\mathcal{Q}_+$, $\mathcal{Q}_F$ in the basis $| K_{-,0} \rangle$, $| K_{+,0}\rangle$, $ | K_{0,-}\rangle$ and $| K_{0,+}\rangle$.
On the other hand, the most general two-particle state can be decomposed as
\begin{eqnarray}
&\vert A_{i,j}(\theta)A_{j,k}(-\theta)\rangle = a\,\vert A_{-1,0}(\theta),A_{0,1}(-\theta)\rangle + b\,\vert A_{-1,0}(\theta),A_{0,-1}(-\theta)\rangle + c\,\vert A_{1,0}(\theta), A_{0,-1}(-\theta)\rangle\nonumber\\
& + d\,\vert A_{1,0}(\theta), A_{0,1}(-\theta)\rangle + e\,\vert A_{0,-1}(\theta), A_{-1,0}(-\theta)\rangle + f\,\vert A_{0,1}(\theta), A_{1,0}(-\theta)\rangle\,\,\,,\label{twoparticletim}
\end{eqnarray}
and we can then compute the action of $(Q_+ \pm i Q_-)$ on each of these two-particle states:
\begin{eqnarray}
&a\,(Q_+ \pm i Q_-)\vert A_{-1,0}(\theta) A_{0,1}(-\theta)\rangle = a\sqrt{m}\left[i(-x\pm x)+(-x^{-1}\pm x^{-1})\right]\left(\begin{array}{c}0\\1\\0\\0\end{array}\right)\otimes\left(\begin{array}{c}0\\0\\0\\1\end{array}\right),\nonumber\\
&b\,(Q_+ \pm i Q_-)\vert A_{-1,0}(\theta),A_{0,-1}(-\theta)\rangle = b\sqrt{m}\left[i(-x\mp x)+(x^{-1}\pm x^{-1})\right]\left(\begin{array}{c}0\\1\\0\\0\end{array}\right)\otimes\left(\begin{array}{c}0\\0\\1\\0\end{array}\right),\nonumber
\end{eqnarray}
\begin{eqnarray}
&c\,(Q_+ \pm i Q_-)\vert A_{1,0}(\theta),A_{0,1}(-\theta)\rangle = c\sqrt{m}\left[i(x\mp x)+(x^{-1}\mp x^{-1})\right]\left(\begin{array}{c}1\\0\\0\\0\end{array}\right)\otimes\left(\begin{array}{c}0\\0\\1\\0\end{array}\right),\nonumber\\
&d\,(Q_+\pm i Q_-)\vert A_{1,0}(\theta),A_{0,1}(-\theta)\rangle = d\sqrt{m}\left[ i(x\pm x)+(-x^{-1}\mp x)\right]\left(\begin{array}{c}1\\0\\0\\0\end{array}\right)\otimes\left(\begin{array}{c}0\\0\\0\\1\end{array}\right)\nonumber\\
&e\,(Q_+ \pm i Q_-)\vert A_{0,-1}(\theta),A_{-1,0}(-\theta)\rangle\nonumber\\
&=e\sqrt{m}\left[(x\mp ix^{-1})\left(\begin{array}{c}0\\0\\1\\0\end{array}\right)\otimes\left(\begin{array}{c}1\\0\\0\\0\end{array}\right)+(-ix^{-1}\pm x)\left(\begin{array}{c}0\\0\\0\\1\end{array}\right)\otimes\left(\begin{array}{c}0\\1\\0\\0\end{array}\right)\right]\nonumber\\
&f\,(Q_+\pm i Q_-)\vert A_{0,1}(\theta),A_{1,0}(-\theta)\rangle\nonumber\\
&=f \sqrt{m}\left[(-x\pm ix^{-1})\left(\begin{array}{c}0\\0\\0\\1\end{array}\right)\otimes\left(\begin{array}{c}0\\1\\0\\0\end{array}\right)+(ix^{-1}\mp x)\left(\begin{array}{c}0\\0\\1\\0\end{array}\right)\otimes\left(\begin{array}{c}1\\0\\0\\0\end{array}\right)\right].\label{expandedtwoparticletim}\nonumber
\end{eqnarray}
By inspecting these conditions, we can see that the equations (\ref{SUSYTIMcondition}) are satisfied by choosing, for the 
first condition, expressed by $(Q_+ + i Q_-)|B \rangle = 0$,
\begin{eqnarray}
b,d=0,\,\,\,\,\,\,e=f,\,\,\,\,\,\,a,c={\rm any \,values},
\end{eqnarray}
and for the second condition, given by $(Q_+ - i Q_-) | B \rangle = 0$,
\begin{eqnarray}
a,c=0,\,\,\,\,\,\,e=-f,\,\,\,\,\,\,b,d={\rm any\,value}.
\end{eqnarray}
In the first case, the boundary states can then be written as
\begin{eqnarray}
&&\vert B^+\rangle = \exp \left\{ \int_{-\infty}^{\infty}d\theta \left[ K^{- +}(\theta)A^\dag_{-1,0}(\theta)A^\dag_{0,1}(-\theta)+K^{+ -}(\theta)A^\dag_{1,0}(\theta)A^\dag_{0,-1}(-\theta)\right.\right.\nonumber\\
&& +\left.\left.K^{0,0}(\theta)\left(A^\dag_{0,-1}(\theta)A^\dag_{-1,0}(-\theta)+A^\dag_{0,1}(\theta)A^\dag_{1,0}(-\theta)\right)\right]\right\}\vert 0\rangle,
\end{eqnarray}
while in the second case 
\begin{eqnarray}
&&\vert B^-\rangle = \exp \left\{ \int_{-\infty}^\infty d\theta \left[ +K^{- -}(\theta)A^\dag_{-1,0}(\theta)A^\dag_{0,-1}(-\theta)+K^{+ +}(\theta)A^\dag_{1,0}(\theta)A^\dag_{0,1}(-\theta)\right.\right.\nonumber\\
&&\left.\left.+K^{0,0}(\theta)\left(A^\dag_{0,-1}(\theta)A^\dag_{-1,0}(-\theta)-A^\dag_{0,1}(\theta)A^\dag_{1,0}(-\theta)\right)\right]\right\}\vert 0\rangle
\,\,\,.
\end{eqnarray}

\vspace{3mm}
\noindent
{\bf Boundary State with Non-Zero Topological Charge}. Let us now consider a boundary state which has a topological charge $z$, such that 
\begin{eqnarray}
Z\vert B\rangle \,=\, z\vert B\rangle.
\end{eqnarray}
In this case, however, the SUSY condition (\ref{susyconstraint}) on $\vert B\rangle$ is incompatible with the spin-1 condition (\ref{integrableboundary}), since
\begin{eqnarray}
(Q_+ \pm i Q_-)(Q_+ \pm i Q_-)\vert B\rangle \,=\, 
(P_+^{(1)} -P_-^{(1)} \pm i z) \vert B\rangle \neq 
(P_+^{(1)} -P_-^{(1)}) \vert B\rangle.
\end{eqnarray}
The generalization of the SUSY condition (\ref{susyconstraint}) that is compatible with (\ref{integrableboundary}) is instead the following \cite{nepomechie}
\begin{eqnarray}
(Q_+ \pm i Q_- +\beta\,\Gamma\,)\vert B\rangle\,=\, 0,\label{susyconstraintz}
\end{eqnarray}
such that
\begin{eqnarray}
(Q_+ \pm i Q_- +\beta\,\Gamma\,)^2\,=\, (P_+^{(1)} -P_-^{(1)})\,\,\,,
\end{eqnarray}
where $\beta^2 = \mp iz$, and $\Gamma$ is the spin reversal operator, which satisfies 
\begin{eqnarray}
\Gamma^2\,=\, 1, \,\,\,\,\{\Gamma,Q_+\} \,=\, \{\Gamma,Q_-\}\,=\, 0.
\end{eqnarray}
The action of $\Gamma$ on two-particle states is given by
\begin{eqnarray}
\Gamma\,\vert A_{i,j}(\theta) A_{j,k}(-\theta)\rangle = \mathcal{Q}_F \otimes\mathcal{Q}_F \,\vert A_{i,j}(\theta) A_{j,k}(-\theta)\rangle.
\end{eqnarray}

Let's now examine what condition eq.\,(\ref{susyconstraintz}) implies for a two-particle state of the form (\ref{twoparticletim}). Repeating a calculation analog to (\ref{expandedtwoparticletim}) we obtain the conditions
\begin{eqnarray}
&a\left[i(-x\pm x)+(-x^{-1}\pm x^{-1})\right]=-b\beta,\,\,\,\,\,\,\,\,\,a\beta=-b\left[i(-x\mp x)+(x^{-1}\pm x^{-1})\right],\nonumber\\
&\nonumber\\
&c\left[i(x\mp x)+(x^{-1}\mp x^{-1})\right]=-d\beta,\,\,\,\,\,\,\,\,\, c\beta=-d\left[i(x\pm x)+(-x^{-1}\mp x)\right],\nonumber\\
&\nonumber\\
&e(x\mp i x^{-1})=-f\beta-f(ix^{-1}\mp x),\,\,\,\,\,\,\,\,\, e(-ix^{-1}\pm x)+e\beta=-f(-x\pm ix^{-1}).\label{conditionsz}
\end{eqnarray}
The only solution to (\ref{conditionsz}) with $\beta\neq 0$ ($z\neq0$), however, is 
\begin{eqnarray}
a=b=c=d=e=f=0.
\end{eqnarray}
Said differently, it is impossible to construct a supersymmetric boundary state out of particle pairs with a non-zero topological charge. This no-go result for the supersymmetric boundary states with a non-zero topological charge can be understood in the following way. A non-zero topological charge signals dominance of the kinks (or anti-kinks) in the state. Such state is not invariant under the kink into anti-kink transformation. However the kinks and anti-kinks are directly related, in the SUSY language, to the bosons and fermions. Therefore nonzero topological charge simply means that $N_b \neq N_f$ which explicitly breaks SUSY.

\begin{figure}[t]
\psfig{figure=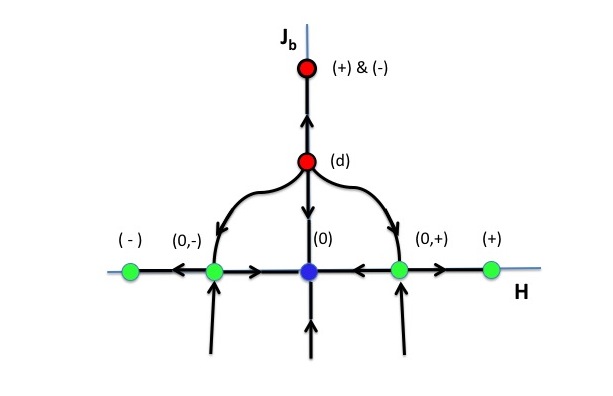,height=6cm,width=9cm}
\caption{{\em Conformal Invariant Boundary States in the Tricritical Ising Model, where $H$ is a Boundary Magnetic Field while $J_b$ is a coupling equivalent to a Temperature Boundary, connected by Renormalization Group flows, according to ref.\cite{AffleckTIM}. The SUSY Boundary States are 
the two upper ones, drawn in red color. }}
\label{TIMBoundaryfigure}
\end{figure}

\vspace{3mm}
\noindent
{\bf Comparison with known boundary solutions}. The Tricritical Ising Model with boundary has been studied in detail in a series of papers, 
both at and away from criticality \cite{Chim,nepomechieTIM,AffleckTIM,CardyBoundary}. Therefore also in this case it is very useful to compare the boundary states identified by the SUSY equations with those already discussed in the literature. As we are going to show, the two classes of boundary states with zero topological charge, defined through $\left(Q_+ \pm i Q_-\right)|B\rangle = 0$ are in one to one correspondence with the SUSY spatial boundary conditions of the TIM studied by Nepomechie in \cite{nepomechieTIM} and there denoted as Ramond (R) and Neveu-Schwartz (NS) boundary conditions. This terminology comes from the general classification of the Conformal Boundary States that the Tricritical Ising Model can have: since this model has six primary operators under the Virasoro algebra, according to the analysis of Cardy \cite{CardyBoundary}, there could be six possible conformal invariant boundary states. At criticality, these six conformal boundary states are expressed as linear combination of Ishibashi states and can be denoted as \cite{Chim,nepomechieTIM,AffleckTIM}: $(-),(+),(0),(- 0),(0+),(d)$. The first three states $(+),(-),(0)$ denote that, at the boundary, the order parameter has been fixed in one of the three vacua of the theory.  The boundary states indicated by $(-0)$ and $(0+)$ mean that, at the boundary,  the vacuum $0$ is degenerate either with the vacuum $-1$ or $+1$. Finally, the boundary state indicated by $(d)$ means that the three vacua are all degenerate at the boundary. Away from criticality, Nepomechie \cite{nepomechieTIM} identified which of these boundary states also respect SUSY: the first one is given by a superposition $(-0) \,\&\, (0+)$ while the second is given by the boundary state $(d)$, and these boundary states were denoted respectively as Neveu-Schwartz and Ramond states. All these boundary conditions, together with the Renormalization Group flows that connect them, are given in Figure \ref{TIMBoundaryfigure}.  

After this brief summary on the boundary states of the Tricritical Ising Model, let's now recall the reflection scattering theory with the boundary 
in space placed at $x=0$: there are six different channels, which can be expressed as  
\begin{eqnarray}
&&  |A_{1,0}(\theta) B_0\rangle_{\rm in}\,\, \,\,\,\,\,=\, R_+(\theta) |A_{1,0}(-\theta) B_0\rangle_{\rm out},\nonumber \\
 && |A_{-1,0}(\theta) B_0\rangle_{\rm in} \,\,\,\,=\, R_-(\theta) |A_{-1,0}(-\theta) B_0\rangle_{\rm out},\\
 & &|A_{0,1}(\theta) B_1\rangle_{\rm in} \,\,\,\,\,\,\,\,=\, P_+(\theta) |A_{0,1}(-\theta) B_{1}\rangle_{\rm out} + V_+(\theta)|A_{0,-1}(-\theta) 
B_{-1}\rangle_{\rm out},\nonumber \\
 & &|A_{0,-1}(\theta) B_{-1}\rangle_{\rm in}  \,=\, P_-(\theta) |A_{0,-1}(-\theta) B_{-1}\rangle_{\rm out} + V_-(\theta)|A_{0,1}(-\theta) B_{1}\rangle_{\rm out},\nonumber
\end{eqnarray}
where $B_a$, with $a \in \{-1,0,1\}$, is the spatial boundary operator associated to each possible vacuum state. In the NS sector the amplitudes fulfill
\begin{eqnarray} 
&&  P_+(\theta) \,=\, P_-(\theta)\, =\, P(\theta),\nonumber\\
&&  R_{\pm}(\theta) \,=\, \frac{1}{2}\left(\cos \zeta/2 \pm i \sinh \theta/2 \right) \,\sigma(\theta - i \zeta) \,\sigma(\theta + i \zeta)\,P(\theta),\\
 && \,V_{\pm} \,\,\,\,\,\,\,\,\,\,\,=\, 0,\nonumber
\end{eqnarray}
where $\sigma(\theta)$ is given in eq.(\ref{sigmafactorTIM}), while in the R sector we have 
\ea {
  P_+(\theta) &= P_-(\theta) \,=\, P(\theta),\nonumber \\
  R_+(\theta) &= R_-(\theta) \,=\, R(\theta),\\
  r V_-(\theta) &= \frac{1}{r} V_+(\theta)\,\,\,\,\,\,\,\,\,,\,\,\,\,\,\, \quad P(\theta) = \frac{ir\cos \zeta/2}{\sinh \theta/2} V_-(\theta) \nonumber.
  }
The parameter $r$ is related to the boundary $g$-factors (see the papers \cite{Chim,nepomechieTIM,AffleckTIM}) for detail) while 
the parameter $\zeta$ is related to a boundary coupling: this parameter rules the energies of the excited states on the boundary, 
related by 
\be 
e_0 \,=\, e_{\pm} +  \cos\zeta \,\,\,.
\ee
Hence at $\zeta = \pi/2$, the boundary state $(0)$ is degenerate either with $(-)$ or with $(+)$. Recalling that the anti-kink of $|A_{ab}(\theta)\rangle$ is $|A_{ba}(\theta)\rangle$, we can again relate the amplitudes $P_{\pm}(\theta)$, $R_{\pm}(\theta)$ and $V_{\pm}(\theta)$ of the boundary scattering processes to amplitudes of the particle pairs. We find
\ea {
  P_+\left(i\frac{\pi}{2} -\theta\right) \,=\, d(\theta), && P_-\left(i\frac{\pi}{2} - \theta\right) \,=\, b(\theta),\\
  R_+\left(i\frac{\pi}{2} -\theta\right) \,=\, f(\theta), && R_-\left(i\frac{\pi}{2} - \theta\right) \,=\, e(\theta),\\
  V_+\left(i\frac{\pi}{2} - \theta\right) \,=\, c(\theta), && V_-\left(i\frac{\pi}{2} - \theta\right) \,=\, a(\theta). 
}
So, we are now in the position to compare our findings with those coming from previous analysis: 
\begin{itemize}
\item the first condition $(Q_+ + i Q_-) | B\rangle =0$ corresponds to the Ramond scattering condition for the value of the parameter $\zeta$ equal to 
$\zeta=\pi/2$. 
\item the other condition $(Q_+ - i Q_-) |B \rangle =0$ corresponds instead to the Neveu-Schwartz scattering condition once again for $\zeta = \pi/2$. 
\end{itemize}
Notice that in both cases we have $\zeta = \pi/2$, which is the only value where there are degeneracy of the energies of the boundary states 
and the maximum symmetry between them, which is after all the requisite to have SUSY. 

\section{Signatures of SUSY in the dynamics out of equilibrium}\label{SUSYsignature}

A theory that is invariant under SUSY permits to find relationships at any time $t$ among the correlation functions of the various fields entering the irreducible multiplets. Let's see how this works considering initially the simple case of the one-point functions:  the quantity we are interested in is 
the expectation value of the superfield $\Phi(x,\theta)$ in the SUSY boundary state
\ea {
  \langle B(t) | \Phi(x,\theta) | B(t) \rangle \,\,\,.
}
If $| B(t) \rangle$ is a SUSY boundary state, combining together the invariance of this state with the transformation law of the various components of the superfield $\Phi$, one expects to find certain constrains on the expectation values of these components. Let's adopt the notation of Section 
\ref{moreond=2}, in which the superfield $\Phi(x, \theta)$ expands as  
\ea {
 \Phi(x,\theta) = \phi(x) + \bar{\theta}\psi(x) + \frac{1}{2}\bar{\theta}\theta F(x)\,\,\,.  
}
A generic supersymmetry transformation can be written as $e^{i\bar{\epsilon}Q}$ and interpreted as a translation operator in the superspace
\ea {
 e^{i\bar{\epsilon}Q} \Phi(x,\theta) e^{-i\bar{\epsilon}Q}\,=\, \Phi(x^{\mu} + \bar{\epsilon}\gamma^{\mu} \theta, \theta + \epsilon).
}
Therefore, under an infinitesimal transformation, we get
\ea {
  \delta_{\epsilon} \Phi = \Phi(x^{\mu}+\bar{\epsilon}\gamma^{\mu} \theta, \theta + \epsilon) - \Phi(x,\theta) = \delta_{\epsilon}\phi(x) + \bar{\theta}\,\delta_{\epsilon}\psi(x) + \frac{1}{2}\bar{\theta}\theta\,\delta_{\epsilon} F(x)+ \mathcal{O}(\epsilon^2),
}
with
\ea { \label{components_transformation}
 \delta_{\epsilon} \phi &= \bar{\epsilon} \psi,\\
 \delta_{\epsilon} \psi &= -i\gamma^{\mu}\epsilon\partial_{\mu}\phi + \epsilon F,\\
 \delta_{\epsilon} F &= -i \bar{\epsilon}\gamma^{\mu} \partial_{\mu} \psi.
}
Using the spinor notation
\ea {
 \epsilon_{\pm} = \epsilon_0 \begin{pmatrix}
   1\\
   \pm i
 \end{pmatrix},\quad\quad
 Q = \begin{pmatrix}
   Q_+\\
   Q_-
 \end{pmatrix},
}
the SUSY boundary state satisfies one of the following conditions  
\ea {
 \bar{\epsilon}_{\pm} Q |B(t)\rangle \,=\, 0\,\,\,.
}
This implies the following identity
\ea {
 \langle B(t)| \Phi(x,\theta)|B(t)\rangle \,=\,  \langle B(t) | e^{i\bar{\epsilon}_{\pm} Q} \Phi(x,\theta) e^{-i\bar{\epsilon}_{\pm} Q} |B(t)\rangle \,\,\,.
}
Expanding for small $\epsilon_0$, this equation gives rise to 
\ea {
 0 \,=\, \langle B(t) |  [ \bar{\epsilon}_{\pm} Q, \Phi(x,\theta) ] |B(t)\rangle \,=\, 
 \langle B(t) |  \left(\bar{\epsilon}_{\pm}Q \Phi(x,\theta)\right) |B(t) \rangle \,=\, \langle B(t)|  \left(\delta_{\epsilon_{\pm}}\Phi(x,\theta)\right) 
 | B(t)\rangle \,\,\,,
}
or in the components
\ea {
 \langle B(t)| \bar{\epsilon}_{\pm}\psi |B(t)\rangle &= 0, \\
 \langle B(t)| -i\gamma^{\mu}\epsilon_{\pm}\partial_{\mu}\phi + \epsilon_{\pm} F  |B(t)\rangle &= 0, \label{one_point_2nd_cnd}\\
 \langle B(t)| \bar{\epsilon}_{\pm}\gamma^{\mu} \partial_{\mu} \psi |B(t)\rangle &= 0. \label{one_point_3rd_cnd}
}
Writing out explicitly the first condition we find
\ea {
 \langle B(t)| \psi_+(x) |B(t)\rangle \,=\, \pm i \langle B(t)| \psi_-(x) |B(t)\rangle,
}
and, since $\psi_{\pm}(x)$ are Majorana fermions, in both cases these equations imply 
\ea {
 \langle B(t)| \psi_{\pm}(x) |B(t)\rangle \,=\, 0 \,\,\,. 
}
The second condition, expressed by eq.~\eqref{one_point_2nd_cnd}, actually gives rise in both cases to two equalities which can be organised 
as  
\begin{eqnarray} 
  \langle B(t) | \partial_- \phi(x) | B(t)\rangle & \,= & \, \langle B(t) | \partial_+ \phi(x) | B(t)\rangle\,\,\,,\\
  \langle B(t) | F(x) | B(t) \rangle &\,= & \,   \pm i \langle B(t) | \partial_- \phi(x) | B(t)\rangle\,\,\,. \label{one_point_F}
\end{eqnarray}
where $\partial_{\pm} = \partial_0 \pm \partial_1$. The last equation implies that the matrix element of the auxiliary field $F$ on the boundary state 
$|B(t) \rangle$ does not follow from the classical equation of motion which would require
\ea {
  F \,=\, - W'(\phi)\,\,\,.
}
For example, for a free massive SUSY theory $W'(\phi) \,=\, m\phi$ which contradicts \eqref{one_point_F}. This implies that dynamics out of 
equilibrium which follows from the quantum quench with SUSY boundary state is governed by field configurations which are far from the classical solutions. 

We can also find exact relations between higher multi-point correlation functions. For the two point function we have
\ea {
 \langle B(t)| \Phi(x_1,\theta_1) \Phi(x_2, \theta_2)| B(t)\rangle \,=\, 
 \langle B(t) | e^{i\bar{\epsilon}_{\pm} Q} \Phi(x_1,\theta_1) \Phi(x_2, \theta_2)e^{-i\bar{\epsilon}_{\pm} Q} | B(t)\rangle \,\,\,.
}
which implies that
\ea {
  0 =& \langle B(t) |  [\bar{\epsilon}_{\pm} Q, \Phi(x_1, \theta_1)\Phi(x_2, \theta_2) ]| B(t)\rangle \nonumber\\ 
  =& \langle B(t) | \delta_{\epsilon_{\pm}}\Phi(x_1, \theta_1)\,\Phi(x_2,\theta_2) + \Phi(x_1, \theta_1)\,\delta_{\epsilon_{\pm}}\Phi(x_2,\theta_2)| B(t)\rangle.\label{two_point_cnd}
}
The product of the superfields can be expanded in Grasmann variables yielding
\ea {
  \Phi(x_1, \theta_1)\,\delta\Phi(x_2, \theta_2) =& \phi(x_1)\,\delta \phi(x_2) +  \phi(x_1)\,\bar{\theta}_2\delta\psi(x_2) + \frac{1}{2}\bar{\theta}_2\theta_2\,\phi(x_1)\,\delta F(x_2) \nonumber\\
  &+ \bar{\theta}_1 \psi(x_1)\,\delta\phi(x_2) + \bar{\theta}_1\psi(x_1)\, \bar{\theta}_2\delta\psi(x_2) + \frac{1}{2} \bar{\theta}_1\psi(x_1)\, \bar{\theta}_2 \theta_2 \delta F(x_2)\nonumber\\
  &+ \frac{1}{2}\bar{\theta}_1\theta_1 F(x_1)\, \delta\phi(x_2) + \frac{1}{2}\bar{\theta}_1\theta_1 F(x_1)\, \theta_2\delta\psi(x_2) \nonumber\\
  &+ \frac{1}{4}\bar{\theta}_1\theta_1 \bar{\theta}_2\theta_2\, F(x_1) \delta F(x_2),
}
and a similar expression for $\delta\Phi(x_1, \theta_1)\,\Phi(x_2,\theta_2)$
\ea {
  \delta \Phi(x_1, \theta_1)\,\Phi(x_2, \theta_2) =& \delta\phi(x_1)\, \phi(x_2) +  \delta\phi(x_1)\,\bar{\theta}_2\psi(x_2) + \frac{1}{2}\bar{\theta}_2\theta_2\,\delta\phi(x_1)\, F(x_2) \nonumber\\
  &+ \bar{\theta}_1 \delta\psi(x_1)\,\phi(x_2) + \bar{\theta}_1\delta\psi(x_1)\, \bar{\theta}_2\psi(x_2) + \frac{1}{2} \bar{\theta}_1 \delta\psi(x_1)\, \bar{\theta}_2 \theta_2 F(x_2)\nonumber\\
  &+ \frac{1}{2}\bar{\theta}_1\theta_1 \delta F(x_1)\,\phi(x_2) + \frac{1}{2}\bar{\theta}_1\theta_1 \delta F(x_1)\, \theta_2\psi(x_2) \nonumber\\
  &+ \frac{1}{4}\bar{\theta}_1\theta_1 \bar{\theta}_2\theta_2\,  \delta F(x_1)\, F(x_2),
}
For the condition~\eqref{two_point_cnd} to be fulfilled all terms with different powers of Grassmann variables must vanish separately. This leads to a number of condition for expectation values of the fields $\phi(x)$, $\psi_{\pm}(x)$ and $F(x)$ out of which we write explicitly just few. From the terms proportional to $\bar{\theta}_1$ and to $\bar{\theta}_1 \bar{\theta}_2\theta_2$ respectively, we find
\ea {
  \langle B(t)| \psi(x_1)\, \delta_{\epsilon_{\pm}}\phi(x_2) + \delta_{\epsilon_{\pm}}\psi(x_1)\, \phi(x_2)|B(t)\rangle &= 0 \label{two_point_1st_cnd}\\
  \langle B(t)| \psi(x_1)\, \delta_{\epsilon_{\pm}} F(x_2) + \delta_{\epsilon_{\pm}}\psi(x_1)\, F(x_2)|B(t)\rangle &=0. \label{two_point_2nd_cnd}
}
The first condition, after using eq.~\eqref{components_transformation} and some reorganization, gives the following two equations 
\ea {
  \langle B(t)| \psi_+(x_1) \left[\psi_+(x_2) \mp i \psi_-(x_2)\right]|B(t)\rangle = \pm\langle B(t)| \left[\mp i\partial_- \phi(x_1) + F(x_1)\right]\phi(x_2)|B(t)\rangle,\\
  \langle B(t)| \psi_-(x_1) \left[ \psi_+(x_2) \mp i \psi_-(x_2)\right]|B(t)\rangle =  -i \langle B(t)| \left[\pm i\partial_+ \phi(x_1) + F(x_1)\right]\phi(x_2)|B(t)\rangle
}
which leads to two relations between the two-point correlation function of the fermionic and the bosonic fields
\ea {
  &\langle B(t)| \left[\psi_+(x_1)\mp i\psi_-(x_1)\right] \left[\psi_+(x_2)  \mp i \psi_-(x_2)\right]|B(t)\rangle = -2i \langle B(t)| \partial_{x_1^0} \phi(x_1)\,\phi(x_2)|B(t)\rangle, \label{two_point1}\\
  &\langle B(t)| \left[\psi_+(x_1) \pm i\psi_-(x_1)\right] \left[\psi_+(x_2)  \mp i \psi_-(x_2)\right]|B(t)\rangle = \pm 2 \langle B(t)|[-i\partial_{x_1^1} \phi(x_1) + F(x_1)]\phi(x_2)|B(t)\rangle, \label{two_point2}
}
Looking instead to eq.~\eqref{two_point_2nd_cnd}), this gives rise to 
\ea {
  \langle B(t)| \psi_+(x_1)\left[\partial_+\psi_+(x_2) \mp i\partial_-\psi_-(x_2) \right]|B(t) \rangle &= -i \langle B(t)| \left(\mp i\partial_-\phi(x_1) + F(x_1) \right)F(x_2)|B(t)\rangle,\\
    \langle B(t)| \psi_-(x_1)\left[\partial_+\psi_+(x_2) \mp i\partial_-\psi_-(x_2) \right]|B(t) \rangle &= \pm \langle B(t)| \left(\pm i\partial_+\phi(x_1) + F(x_1) \right)F(x_2)|B(t)\rangle,
  }
which can also be written as  
\ea {
  \langle B(t)| \left[\psi_+(x_1) \pm i \psi_-(x_1)\right]\left[\partial_+\psi_+(x_2) \mp i\partial_-\psi_-(x_2) \right]|B(t) \rangle &= \mp 2\langle B(t)| \partial_{x_1^0}\phi(x_1) F(x_2)|B(t)\rangle, \label{two_point3}\\
    \langle B(t)| \left[\psi_-(x_1) \mp i \psi_-(x_1)\right]\left[\partial_+\psi_+(x_2) \mp i\partial_-\psi_-(x_2) \right]|B(t) \rangle &= -2i \langle B(t)| \left(\pm i\partial_{x_1^1}\phi(x_1) + F(x_1) \right)F(x_2)|B(t)\rangle, \label{two_point4}
}
Similarly to the case of the two-point correlation functions, other exact relations can be easily derived for multi-point correlators sandwiched between the SUSY boundary state $|B(t)\rangle$.

\vspace{3mm}
\noindent
{\bf Tricritical Ising Model.} The Tricritical Ising Model is the simplest SUSY model for which the relations above apply. Its Neveu-Schwartz superfield 
is made of the energy density $\epsilon(x)$, the fermion field $\psi(x)$ and the vacancy density $t(x)$ 
\be \label{TIM_superfield22}
\Phi(x,\theta) \,=\,\epsilon(x) + \bar\theta \psi(x) + \frac{1}{2} \bar\theta \theta t(x) \,\,\,,
\ee
and therefore for quench processes induced by SUSY boundary states we have exact and non-perturbative relations which involve the 
correlation functions of these fields.

\vspace{3mm}
\noindent
{\bf Sine-Gordon Model.} We have seen in Section \ref{SGBoundary} that the fixed boundary conditions discussed by Goshal and Zamolodchikov 
\cite{GZ} provide SUSY initial states that satisfy the conditions $(W_+ \pm iW_-)\vert B\rangle=0$. Such states are invariant under the infinitesimal transformation 
\begin{eqnarray}
\delta_{\epsilon_{\pm}}\vert B\rangle=\vert B\rangle,\,\,\,\,\,\,\delta_{\epsilon_{\pm}}=\exp\left[\epsilon_0\left(W_+\pm iW_-\right)\right].
\end{eqnarray}
and therefore we have such an identity 
\begin{eqnarray}
\langle B\vert \delta_{\epsilon_{\pm}}\, \mathcal{O}^n_a(x) \,\delta_{\epsilon_{\pm}}^\dag\vert B\rangle\,=\, \langle B\vert \mathcal{O}^n_a(x)\vert B\rangle.\label{uou}
\end{eqnarray}
for the correlation functions involving the soliton-creating operators $\mathcal{O}^n_a(x)$. To write explicitly the consequences of this identity, one would need to use the commutation relations between the charges $G_{\pm}, \bar{G}_{\pm}$ and the operators $\mathcal{O}_a^n(x)$ \cite{lukyanovzamolodchikov}. These commutation relations are in general nontrivial and involve the operators $\mathcal{O}_{a^\prime}^{n\pm 2}(x)$, with a shifted parameter $a^\prime$, and their descendants. A particularly simple set of commutation relations is found when the soliton-creating operators are of the form $\mathcal{O}^n_{a_n}(x)$, where
\begin{eqnarray}
a_n=\frac{n}{\beta}\sqrt{\frac{\pi}{2}}+\frac{\beta}{2\sqrt{8\pi}}.
\end{eqnarray}
In this case the commutation relations are given by \cite{lukyanovzamolodchikov}
\begin{eqnarray*}
  G_+ \mathcal{O}_{a_{n}}^n (x) - (-i)^n \mathcal{O}_{a_{n}}^n (x) G_+ &=& \frac{2\pi i}{N_Q} (-i)^{n/2} \mathcal{O}_{a_{n+2}}^{n+2} (x), \\
  G_- \mathcal{O}_{a_{n}}^n (x) - (-i)^n \mathcal{O}_{a_{n}}^n (x) G_- &=& \frac{2\pi i}{N_Q} (-i)^{n/2} \mathcal{O}_{a_{n-2}}^{n-2} (x), \\
  \bar{G}_+ \mathcal{O}_{a_{n}}^n (x) - (-i)^n \mathcal{O}_{a_{n}}^n (x) \bar{G}_+ &=& \frac{2\pi i}{N_Q} (-i)^{n/2} \mathcal{O}_{a_{n-2}}^{n+2} (x), \\
  \bar{G}_- \mathcal{O}_{a_{n}}^n (x) - (-i)^n \mathcal{O}_{a_{n}}^n (x) \bar{G}_- &=& \frac{2\pi i}{N_Q} (-i)^{n/2} \mathcal{O}_{a_{n+2}}^{n-2} (x),
\end{eqnarray*}
such that, building the two SUSY charges $W_+ = Q_+ + Q_-$ and $W_- = i (\bar{Q}_+ - \bar{Q}_-)$, we find
\begin{eqnarray}
  W_+ \mathcal{O}_{a_{n}}^n (x) - (-i)^n \mathcal{O}_{a_{n}}^n (x) W_+ &=& \frac{2\pi i}{N_Q} (-i)^{n/2} \left( \mathcal{O}_{a_{n+2}}^{n+2} (x) + \mathcal{O}_{a_{n-2}}^{n-2} (x)\right),\\
  W_-\mathcal{O}_{a_{n}}^n (x) - (-i)^n \mathcal{O}_{a_{n}}^n (x) W_- &=& i \frac{2\pi i}{N_Q} (-i)^{n/2} \left( \mathcal{O}_{a_{n-2}}^{n+2} (x) - \mathcal{O}_{a_{n+2}}^{n-2} (x)\right).
\end{eqnarray}
Eq. (\ref{uou}) then implies the following nontrivial relations 
\begin{eqnarray}
\langle B(t)\vert \mathcal{O}_{a_{n+2}}^{n+2}+\mathcal{O}_{a_{n-2}}^{n-2} \pm \mathcal{O}_{a_{n-2}}^{n+2} 
\mp \mathcal{O}_{a_{n+2}}^{n-2}\vert B(t)\rangle\,=\, 0.\label{boundary_equation}
\end{eqnarray}
between the one-point functions of the soliton-creating operators. These are the simplest relations implied by SUSY. In general, if the operator in Eq. (\ref{uou}) is not of the form $\mathcal{O}^n_{a_n}(x)$, there will be instead more articulated relations (that will not be written here) which also involve descendant operators. Relations between two-point and higher-point correlation functions can be found similarly. 

\section{The Generalized Gibbs Ensemble in presence of SUSY}\label{SUSYGGE}

In this section we discuss some special features of an integrable SUSY theory a long time after a quantum quench. Understanding how an integrable model equilibrates is a subject of ongoing research, see for instance \cite{Weiss,Deutsch,CC,SilvaReview,IC,Berges,Rigol,FM,EMP,CEF,GGErecent,Bertin}. It is generally accepted that after a quantum quench, integrable theories do not equilibrate into an ordinary thermal state rather that the presence of nontrivial conserved charges leads to infinite-time dynamics governed by a generalized Gibbs ensemble (GGE) \cite{Rigol}. The ensemble averages are computed in this case using the density matrix
\begin{eqnarray}
  \rho_{\rm GGE} \,=\, \frac{1}{\mathcal{Z}}\exp\left( -\sum_{j=1}^{\infty} \beta_j I_j\right) \,\,\,, 
\end{eqnarray}
where $I_j$ is the conserved charge and $\beta_j$ is a chemical potential associated to the $j$-th charge. These chemical potentials are found, in principle, by requiring that the expectation value of the conserved charges in the initial state is equal to the expectation value in the GGE ensemble
\begin{eqnarray}
  \langle B| I_j |B\rangle = {\rm tr}\left(I_j \rho_{\rm GGE} \right) \,\,\,\, ,\,\,\, \quad j=1,\dots.
\end{eqnarray}
Which conserved charges have to be used to construct the GGE and whether they have to be strictly local or not, and moreover if there is a set of conserved charges better than others,  are all questions still debated in the literature \cite{Rigol,FM,EMP,CEF,GGErecent}. 

For the SUSY integrable theories there are two types of conserved charges: (a) the first type consists of the usual bosonic conserved charges which, strictly speaking, are those responsible for the integrability of the theory;  (b) the second type are the fermionic SUSY charges. The main difference between the bosonic and fermionic conserved charges is that the later quantities are not extensive. One way to see that the SUSY charges are not extensive is the following: since the SUSY charges are expressed in terms of the space integral of the zero component of the supercurrent  -- see 
eqs.\,(\ref{supercurrent}) and (\ref{supercharge}) -- going to Fourier transform we have 
\be 
Q_{\alpha} \,=\,\int dx \,J^0_{\alpha}(x) \,=\,\hat J^0_{\alpha}(0) \,\,\,,  
\ee
where $\hat J^0_{\alpha}(k)$ is the Fourier transform of the supercurrent. To be an expensive quantity, $Q_\alpha$ must then be proportional to $L$ where $L$ is the volume of the system: this means that the zero-mode of the supercurrent must be macroscopically populated, which is impossible since $\hat J_\alpha(k)$ is a fermionic operator. 

A proposal to overcome this problem and to open the possibility to define operators associated to the SUSY charges which are extensive was done in \cite{Kapusta}. It makes use of the Clifford algebra and Clifford numbers: a Clifford number $c$ is fermionic in the sense that it commutes with all bosonic operators $B$ and anti-commutes with all other fermionic operators $F$, but whose square is unity 
\be
[c,B] \,=\,\{c,F\} \,=\,0 
\hspace{5mm} ; 
\hspace{5mm} 
\{c(p),c(q)\} = 2 \delta(p-q) \,\,\,\,\,,\,\,\,\,\, c^{\dagger}(p) = c(p) \,\,\,. 
\ee 
One can now define two new operators, which this time are bosonic  
\be
\tilde Q_\pm \,=\, i \int dp \,c(p)\, J^0_\pm(-p) \,\,\,, 
\ee
with the properties 
\be 
\tilde Q_{\pm}^\dagger = \tilde Q_\pm
\hspace{5mm}
,
\hspace{5mm}
[\tilde Q_\pm, H] \,=\, 0
\hspace{5mm}
,
\hspace{5mm}
\tilde Q^2_+ + \tilde Q^2_- \,=\, 2 H \,\,\,. 
\ee 
These new charges are now extensive since there is no obstruction to have a macroscopically populated mode made of 
\textquotedblleft {\em Cooper pairs} \textquotedblright  of the fermionic fields $c(k)$ and $\hat J(-k)$, analogously to what happens for the ground state of superconductor. These $\tilde Q_{\pm}$ are then the charges that can be used in the SUSY Generalised Gibbs Ensemble (SGGE) average, together with the rest of the other bosonic charges $I_j$ 
\begin{eqnarray}
  \rho_{\rm SGGE}\,=\, \frac{1}{\mathcal{Z}}\exp\left( - \mu_+ \tilde Q_+ - \mu_- \tilde Q_- - \sum_{j=1}^{\infty} \beta_j I_j\right) \,\,\,. 
\end{eqnarray}
This is the more general formulation of the equilibrium ensemble of an integrable SUSY theories for a generic boundary state. Few comments: notice 
that the chemical potential associated to the Hamiltonian (alias the inverse of the temperature) cannot be consider independent from the chemical potentials $\mu_{\pm}$ of the super-charges; moreover, another important difference with respect to the usual GGE density matrix in purely bosonic theories is that even though the supercharges commute with the Hamiltonian, they are not simultaneously diagonal with it. We will not enter further tthe discussion of these details here simply because for the SUSY boundary states there is a drastic simplification: indeed, from the hermiticity of the SUSY charges it follows that $\langle B|Q_+|B\rangle$ and $\langle B|Q_-|B\rangle$ are both real numbers; however if $|B\rangle$ is a SUSY initial state which satisfies one of these conditions, $(Q_+ \pm i Q_-)|B\rangle = 0$, then
\begin{eqnarray}
  \langle B|Q_+|B\rangle \,=\, \mp i\langle B|Q_-|B\rangle.
\end{eqnarray}
which can be fulfilled only if both expectation values are zero. Therefore for the SUSY boundary states the expectation values of the SUSY charges are exactly zero. As a consequence, the SUSY charges do not enter the GGE, as the chemical potentials $\mu_\pm$ associated to them are also identically zero, and therefore for these SUSY states the density matrix of equilibrium is made of only the bosonic conserved charges.

\section{Conclusions and Perspectives}\label{conclusions}
In this paper we have considered the dynamics out of equilibrium which follows a quantum quench in a field theory invariant under supersymmetric transformations. For simplicity, we have restricted our attention to $N=1$ SUSY models. The special properties of the SUSY 
theories, in particular the direct link that exists between the SUSY generator and the Hamiltonian, introduce distinguished signatures in the dynamics out of equilibrium of these theories, such as a spontaneous symmetry breaking but without neither the appearance of Goldstino mode nor the splitting of the degeneracy of bosonic and fermionic particles. Moreover, as argued in the paper, at least in two-dimensional integrable cases one can still have a perfect matching between the mode occupations of the bosons and fermions. All these facts imply that, if the initial boundary state $| B \rangle$ of the quench process is properly chosen, then there could be a perfect SUSY formulation for the entire dynamics out of equilibrium, with a consequent series of identities among the correlation functions of various fermionic and bosonic fields. In this paper we have studied the 
consequences coming from those initial states which are solution of one of these conditions $(Q_+ \pm i Q_-) | B \rangle =0$. We have seen that 
these are sufficient conditions to have a series of remarkable consequences, among which a SUSY time evolution and a final GGE formulation which 
is the closest as possible to the GGE of the bosonic theories, since the expectation values of both charges on these special states $| B \rangle$ vanish, $\langle B | Q_{\pm} | B \rangle = 0$.

\begin{figure}[t]
\psfig{figure=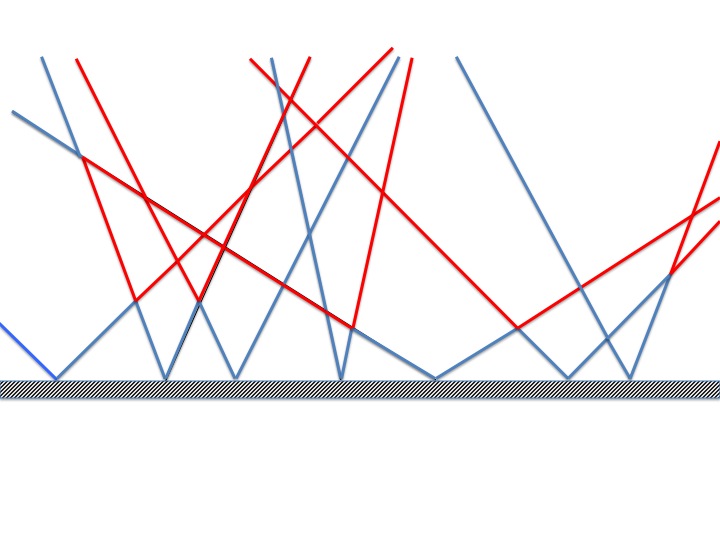,height=5cm,width=9cm}
\caption{{\em A boundary state initially made of only bosonic excitations (blue lines) which therefore explicitly breaks SUSY. However  
its time evolution may give rise dynamically to a mixture of bosonic and fermionic particles (red lines) which may asymptotically restore 
the SUSY of the theory.}}
\label{SUSYRESTORATIONFIGURE}
\end{figure}

In this paper, to illustrate various aspects of our analysis we have chosen as significant examples two models which may find an experimental realisation in the next future: the first is given by the Sine-Gordon model, the second by the Tricritical Ising Model. In both cases, the underlying SUSY of these models has a non-trivial representation in terms of their topological excitations and rules as well their scattering amplitudes. In the case of the Sine-Gordon model, its recent experimental implementation \cite{Schweigler} opens new perspectives on the possibility to perform direct experiments for probing the SUSY physics underling such a model. Moreover, the recent theoretical results on the quench dynamics of such a model \cite{Bertin} 
may be useful to check some of our findings on correlation functions out of equilibrium. Concerning the quantum field theory associated to the Tricritical Ising Model, it describes the scaling region of the Blume-Capel spin chain of spin $1$ systems and there are various magnetic compounds, such as $\displaystyle{Ni (C_2 H_8 N_2) Ni (CN)_4}$ or similar, which may be used to realise such a class of universality and test its SUSY.  It is also worth to mention a recent proposal to realise the SUSY of this class of universality in terms of strongly interacting chain of Majorana zero modes \cite{Affleckstrongly}. 

It must be stressed that this paper has addressed only few questions of SUSY out of equilibrium but has left open many others: for instance we have identified a class of initial states for which there is a SUSY time evolution and, a fortiori, an asymptotic equilibrium situation that is also supersymmetric. But it remains to be seen whether one could have an asymptotic equilibrium SUSY situation also starting from an initial state that explicitly breaks SUSY, alias if it could be a dynamical restoration of SUSY if it was initially broken. This is an an important point which will be extremely interesting to investigate in the future, both for the integrable and non-integrable theories. In both theories, pairs of bosons can be dynamically converted into pairs of fermions and viceversa 
\be
 b \, b \Leftrightarrow f\,f \,\,\,, 
 \ee
and this continuous interchange between the two type of particles may lead to a stationary state where there is a perfect balance between the two species, with a consequent SUSY invariance in the asymptotic configuration of the system, see Figure \ref{SUSYRESTORATIONFIGURE}. Another compelling aspects that deserves a full analysis is the dynamics out of equilibrium for theories with extended SUSY, in particular $N=2$, in view of the robust analytic structure of these theories and their importance in a number of important branches of theoretical physics. 

\newpage
\vspace{1cm}
\begin{flushleft}\large
\textbf{Acknowledgements}
\end{flushleft}
We would like to thank A. Bastianello, P. Calabrese and R. Konik for interesting discussions. One of us (GM) thanks M. Bertolini, S. Cecotti and G. Martinelli for useful conversations, while MP is grateful to T. Fokkema, K. Schoutens, K. Skenderis and P. Su{\l}kowski for interesting comments. 
This work acknowledges the IRSES grant FP7-PEOPLE-2011-IRSES QICFT 295234. This work was supported by ERC under the Starting Grant n.279391. MP would like to acknowledge hospitality and support from the Galileo Galilei Institute, where initial part of this work was carried out during the programs ``Holographic Methods for Strongly Coupled Systems'' and ``Statistical Mechanics, Integrability and Combinatorics''.

\vspace{3mm}




\begin{thebibliography}{99}

\bibitem{Weiss} 
T. Kinoshita, T. Wenger, D. S. Weiss, Nature 440, 900 (2006); M. Greiner, O. Mandel, T. W. H\"ansch, and I. Bloch, Nature 419 51 (2002); S. Hofferberth, I. Lesanovsky, B. Fischer, T. Schumm, and J. Schmiedmayer, Nature 449, 324 (2007).
\bibitem{Deutsch} J.M. Deutsch, Phys. Rev. A 43, 2046 (1991); M. Srednicki, Phys. Rev. E 50, 888 (1994).
\bibitem{CC} P. Calabrese and J. Cardy, J. Stat. Mech. P06008 (2007)
\bibitem{SilvaReview} A. Polkovnikov, K. Sengupta, A. Silva and M. Vengalattore,
Rev. Mod. Phys. \textbf{83}, 863-883  (2011) and references therein. 
\bibitem{IC} A. Iucci and M. A. Cazalilla, Phys. Rev. A 80, 063619 (2009); T. Barthel, U. Schollw\"ock, Phys.Rev.Lett.100,100601 (2008); S. R. Manmana, S. Wessel, R. M. Noack, A. Muramatsu, Phys. Rev. B 79, 155104 (2009); D. Rossini, A. Silva, G. Mussardo, G. Santoro, Phys.
Rev. Lett. 102, 127204 (2009); D. Rossini, S. Suzuki, G. Mussardo, G. E. Santoro, A. Silva, Phys. Rev. B 82, 144302 (2010)
\bibitem{Berges} J. Berges, S. Borsanyi, C. Wetterich, Phys.Rev.Lett. 93 (2004) 142002; M. Kollar, F. A. Wolf, M. Eckstein, Phys. Rev. B 84, 054304 (2011); M. C. Ban\~uls, J. I. Cirac, and M. B. Hastings, Phys. Rev. Lett. 106, 050405 (2011); G. Biroli, C. Kollath, and A.M. L\"auchli, Phys. Rev. Lett. 105, 250401 (2010); G. P. Brandino, A. De Luca, R.M. Konik, G. Mussardo, Phys. Rev. B 85, 214435; G. Mazza and M. Fabrizio, 
Phys. Rev. B 86, 184303 (2012); T. Caneva, E. Canovi, D. Rossini, G. E. Santoro, A. Silva, J. Stat. Mech. (2011) P07015; 
D. Iyer, N. Andrei, Phys. Rev. Lett. 109, 115304 (2012); D. Iyer, H. Guan, N. Andrei, Phys. Rev. A 87, 053628 (2013); 
J.-S. Caux, R. M. Konik, Phys. Rev. Lett. 109, 175301 (2012); F. H. L. Essler, S. Evangelisti, M. Fagotti, Phys. Rev. Lett. 109, 247206 (2012); 
J.-S. Caux and F. H. L. Essler, Phys. Rev. Lett. 110, 257203 (2013); M. Fagotti, M. Collura, F. H. L. Essler, and P. Calabrese,
Phys. Rev. B 89, 125101 (2014). 
\bibitem{Rigol}
M. Rigol, V. Dunjko, V. Yurovsky, and M. Olshanii, Phys. Rev. Lett. 98, 050405 (2007); 
M.~Rigol, V.~Dunjko, M.~Olshanii, 
Nature {\bf 452}, 854 (2008). 
\bibitem{FM} D. Fioretto, G. Mussardo, New J. Phys. 12, 055015 (2010); S. Sotiriadis, D. Fioretto, and G. Mussardo, J. Stat. Mech. (2012) P02017;  
G. Mussardo, Phys. Rev. Lett. 111, 100401 (2013).
\bibitem{EMP} F.H.L. Essler, G. Mussardo and M. Panfil, Phys. Rev. A 91, 051602 (2015). 
\bibitem{CEF}  P. Calabrese, F.H.L. Essler, and M. Fagotti, Phys. Rev. Lett. 106, 227203 (2011); 
J. Stat. Mech. P07016 (2012); J. Stat. Mech. P07022 (2012).
\bibitem{GGErecent} B. Wouters, J. De Nardis, M. Brockmann, D. Fioretto, M. Rigol
and J.-S. Caux, Phys. Rev. Lett. 113, 117202 (2014); M. Brockmann, B. Wouters, D. Fioretto, J. De Nardis, R. Vlijm and J.-S. Caux, J. Stat. Mech. P12009 (2014); B. Pozsgay, M. Mestyan, M. A. Werner, M. Kormos, G. Zarand and G. Takacs, Phys. Rev. Lett. 113, 117203 (2014); 
M. Mestyan, B. Pozsgay,G. Takacs and M.A.Werner, J.Stat. Mech. P04001 (2015); E. Ilievski, J. De Nardis, B. Wouters, J.-S. Caux F. H. L. Essler and T. Prosen, arXiv:1507.02993. 
\bibitem{Bertin} B. Bertini, D. Schuricht, F. H. L. Essler, J. Stat. Mech. (2014) P10035. 
\bibitem{GMbook} G. Mussardo, {\em Statistical Field Theory}. Oxford Univ. Press, Oxford (2010), and references therein.
\bibitem{Tsvelikbook} A. M. Tsvelik, {\em Quantum Field Theory in Condensed Matter Physics}, Cambridge Univ. Press (2003).    
\bibitem{KMT} M. Kormos, G. Mussardo, A. Trombettoni, Phys.Rev.A 81, 043606 (2010); Phys.Rev.Lett. 103, 210404 (2009). 
\bibitem{WZ} J. Wess, B. Zumino, Nucl. Phys., B70 (1974), 39. 
\bibitem{Sonhius} M.F. Sonhius, Phys. Rept. 128 (1985) 39-204. 
\bibitem{Weinberg} S. Weinberg, {\em The Quantum Theory of Fields, Volume 3: Supersymmetry}, Cambridge Univ. Press, 2005. 
\bibitem{Polchinski} J. Polchinski, {\em String Theory Vol. II: Superstring Theory and Beyond}, Cambridge University Press, 1998. 
\bibitem{mirror} K. Hori et al. {\em Mirror symmetry}, Clay mathematics monography, American Mathematical Society 2003. 
\bibitem{OliveWitten} E. Witten and D. Olive, Phys. Lett. B 76 B (1978), 97.
\bibitem{Kareljan} B. Bauer, L. Huijse, E. Berg, M. Troyer, K. Schoutens, Phys. Rev. B 87, 165145 (2013); 
L. Huijse, D. Mehta, N. Moran, K. Schoutens, J. Vala, New J. Phys. 14, 073002 (2012); 
L. Huijse, N. Moran, J. Vala, K. Schoutens, Phys. Rev. B 84, 115124 (2011); 
L. Huijse, K. Schoutens, Adv. Theor. Math. Phys. 14.2 (2010), 643; P. Fendley, B. Nienhuis, K. Schoutens, J.Phys. A 36:12399, 2003; 
P. Fendley, K. Schoutens, J. de Boer, Phys.Rev.Lett. 90 (2003) 120402. 
\bibitem{devega} J. Baacke, D. Cormier, H.J. de Vega, K. Heitmann, Nucl. Phys. B 649 [FS] (2003), 415. 
\bibitem{3dsusy}  L.Y. Hung, M. Smolkin, E. Sorkin,  JHEP 1312 (2013) 022.  
\bibitem{KWSmirnov} M. Karowski, P. Weisz: Nucl. Phys. B 139 (1978) 455; 
F.A. Smirnov, {\em Form factors in completely integrable models of quantum field theory},
Adv. Series in Math. Phys. 14. (World Scientific, Singapore, 1992). 
\bibitem{1998_Mussardo_NPB_532}
G. Mussardo Nucl. Phys. B 532 (1998), 529. 
\bibitem{Goldstino} A. Salam, J. Strathdee, Phys. Lett., 49 B (1974), 465; J. Iliopoulos, B. Zumino
Nucl. Phys., B76 (1974), 310; D.V. Volkov, V.P. Akulov, Phys. Lett. B, 46 (1973), 109. 
\bibitem{1978_Das_PRD_12}
A. Das A and M. Kaku, Phys. Rev. D 18 (1978), 12. 
\bibitem{Girardello} L. Girardello, M.T. Grisaru, P. Salomonson, Nucl. Phys. B 178, 331 (1981).
\bibitem{SSG} S. Sengupta, P. Majumdar, Phys. Rev.  D 33 (1986), 3138; C. Ahn, Nucl. Phys. B 354 (1991), 57; 
S. Shankar, E. Witten, Phys. Rev.  D 17 (1978), 2134; A.M. Tsvelik, Sov. J. Nucl. Phys. 47 (1988), 172; A. Hegedus, F. Ravanini, 
J. Suzuki, Nucl. Phys. B 763 [FS] (2007), 330; Z. Bajnok, C. Dunning, L. Palla, G. Takacs and F. Wagner, 
Nucl. Phys. B 679 [FS] (2004), 521; M. Sakagami, Nucl. Phys. B 207 (1982), 430; G. Mussardo, arXiv:1508.05975. 

\bibitem{1984_Matsumoto_PRD_29} H. Matsumoto, M. Nakahara, Y. Nakano and H. Umezawa, 
Phys. Rev. D 29 (1984), 12. 

\bibitem{Dasbook} A. Das, {\em Finite Temperature Field Theory}, World Scientific, Singapore 1977. 

\bibitem{SUSYschoutens} K. Schoutens, Nucl. Phys. B344 (1990) 665. 

\bibitem{lukyanovzamolodchikov} S. L. Lukyanov and A. B. Zamolodchikov, Nucl. Phys. B 607 (2001) 437.
\bibitem{DenisAndre} D. Bernard and A. LeClair, Nucl. Phys. B, 340 (1990), 721. 

\bibitem{ZamTIM} A.B. Zamolodchikov, 
{\em Fractional spin integrals of motion in perturbed conformal field theory}, 
in Beijing 1989, Proceedings, Fields, strings and quantum gravity. 
\bibitem{ZamRSOS} Al.B. Zamolodchikov, Nucl. Phys.  B 358 (1991) 497. 
\bibitem{Qiu} 
D. Friedan, Z.A. Qiu, S. H. Shenker, Phys.Lett. B151 (1985) 37;  
Z.A. Qiu, Nucl.Phys. B270 (1986) 205-234. 
\bibitem{MSS} G. Mussardo, G. Sotkov, M. Stanishkov,  Phys. Lett. B 195 (1987) 397;  
Nucl. Phys. B 305 (1988) 69. 
\bibitem{Martinec} 	
D.A. Kastor, E.J. Martinec, S.H. Shenker,  Nucl. Phys. B 316 (1989) 590. 




\bibitem{ZamTBA} Al. B. Zamolodchikov, Nucl. Phys. B 342 (1990), 695. 
\bibitem{TBASUSY}M. Moriconi, K. Schoutens, Nucl. Phys. B 464 [FS] (1996), 472; 
C. Ahn, Nucl. Phys. B 422 (1994) 449. 
\bibitem{GZ} S. Ghoshal and A. Zamolodchikov, Int. J. Mod. Phys. A9 (1994) 3841. 
\bibitem{nepomechie} N.P. Warner, Nucl.Phys. B450 (1995) 663;  
M. Moriconi, K. Schoutens,  Nucl.Phys. B487 (1997) 756;  
C. Ahn, R. I. Nepomechie, Nucl.Phys. B586 (2000) 611-640 
C. Ahn, R. I. Nepomechie, Nucl.Phys. B594 (2001) 660; 
R. I. Nepomechie, Phys.Lett. B509 (2001) 183;  
Phys. Lett. B 516 (2001) 161; J. Phys. A 34 (2001) 6509. 
\bibitem{Chim} L. Chim, Int. J. Mod. Phys. A 11 (1996) 4491.  
\bibitem{nepomechieTIM} 	  
R.I. Nepomechie, C. Ahn, Nucl. Phys. B 647 (2002) 433; 
R.I. Nepomechie, Int.J.Mod.Phys. A 17 (2002) 3809; 
\bibitem{AffleckTIM} I. Affleck, J. Phys. A: Math. Gen. 33 (2000) 6473-6479	
\bibitem{CardyBoundary} J.L. Cardy,  Nucl. Phys. B 324 (1989) 581; Nucl. Phys. B 275 (1986) 200. 
\bibitem{2003_Kratzer_AP_308} K. Kratzert, Ann. Phys. 308 (2003), 285. 
\bibitem{1981_Witten_NPB_185}
E. Witten, 
Nucl. Phys. B 185 (1981), 513. 


\bibitem{Kapusta}
  J.~Kapusta, S.~Pratt and V.~Visnijc
  Phys Rev. D 28 (1983), 3093. 
  



\bibitem{Schweigler}
T. Schweigler et al. \textit{On solving the quantum many-body problem}, 
arXiv:1505.03126 



\bibitem{Affleckstrongly} 
A. Rahmani, X. Zhu, M. Franz, I. Affleck, Phys. Rev. Lett. 115, 166401 (2015). 

\end{thebibliography}
\end{document}